\documentclass[preprint,review,12pt,sort&compress]{elsarticle}
\usepackage[margin=1.18 in]{geometry}
\usepackage{multirow}

 \newcommand{\bc}{\begin{center}}
 \newcommand{\ec}{\end{center}}
                   \newcommand{\bfr}{\begin{flushright}}
                   \newcommand{\efr}{\end{flushright}}
   
     \newcommand{\be}{\begin{enumerate}}
     \newcommand{\ee}{\end{enumerate}}
        \newcommand{\bi}{\begin{itemize}}
        \newcommand{\ei}{\end{itemize}}
            \newcommand{\bd}{\begin{description}}
            \newcommand{\ed}{\end{description}}
                \newcommand{\beq}{\begin{equation}}
                \newcommand{\eeq}{\end{equation}}
                  \newcommand{\bea}{\begin{eqnarray}}
                  \newcommand{\eea}{\end{eqnarray}}

      \newcommand{\bfi}{\begin{figure}}
      \newcommand{\efi}{\end{figure}}
\newcommand{\bay}{\begin{array}{l}}
\newcommand{\eay}{\end{array}}
            
\usepackage{amsmath}
\usepackage{breqn}
\usepackage{textcomp}
\usepackage{float}
\usepackage{booktabs}
\usepackage{graphicx}
\usepackage{hyperref}
\usepackage[dvipsnames]{xcolor}
\usepackage{enumerate}
\usepackage{enumitem}

\usepackage{amssymb}

\begin{document}

\begin{frontmatter}



\title{An Isogeometric Framework for the Modeling of Curvilinear Anisotropic Media}

\author[label1]{Kenta Suzuki}
\author[label3]{Sean E. Phenisee}
\author[label4]{Marco Salviato\corref{cor1}}
 \address{\textsuperscript{1}William E. Boeing Department of Aeronautics and Astronautics, University of Washington, Seattle, WA 98195, USA}

\cortext[cor1]{Corresponding Author, \ead{salviato@aa.washington.edu}}

\begin{abstract}
\linespread{1}\selectfont
The advent of multi-material additive manufacturing and automated composite manufacturing has enabled the design of structures featuring complex curvilinear anisotropy. To take advantage of the new design space, efficient computational approaches are quintessential. In this study, we explored a new NURBS-based Isogeometric Analysis (IGA) framework for the simulation of curvilinear fiber composites and we compared it to standard Finite Element Analysis (FEA). A plate featuring a semi-circular notch under tensile loading with different fiber configurations served as a case study. 

\noindent We showed that, thanks to the exact geometric representation and the enriched continuity between elements, NURBS-based IGA outperforms classical FEA in terms of computational efficiency, time-consumption, and estimation quality of field variables for same number of degrees-of-freedom. To further demonstrate the use of the IGA framework, we performed 
optimization studies aimed at identifying the fiber paths minimizing stress concentration and Tsai-Wu failure index. The model showed that curvilinear anisotropy can be effectively harnessed to reduce the stress concentration of up to 82\% compared to unidirectional composites without affecting the overall plate stiffness significantly.

\end{abstract}
\begin{keyword}
Curvilinear Anisotropy \sep Isogeometric Analysis \sep Finite Element Analysis \sep Fiber optimization \sep Failure criterion
\end{keyword}
\end{frontmatter}



\section{Introduction}

Fiber reinforced composites find broad applications in large and complex primary structures for e.g. aerospace, defence, and automotive thanks to superior specific stiffness, strength, and fatigue properties compared to most metals \cite{Das2019PreparationDO,Harris2002,Othman2018}. 
In recent years, the advent of e.g. Automated Fiber Placement (AFP) and Additive Manufacturing (AM) has widen the range of design space of composite materials, and has broaden the ability to design highly customized complex parts 
with great quality~\cite{dirk2012engineering,CROFT2011484,GAO201565}. 

The first example of the use of curvilinear anisotropy to optimize structural behavior can be found in the 1950s in the pioneering work of Mansfield ~\cite{Mansfield1953} who introduced the concept of ``\textit{neutral hole}". By adding a reinforcement along the boundary of a notch, Mansfield was able to find shapes leading to stress distributions elastically equivalent to the ones of uncut structures. However, these configurations were quite limited because the curvilinear reinforcement could only be applied along the boundary.

A step forward is represented by the work of Hyer and Charette which was the first introducing Curvilinear Transverse Isotropy (CTI) in the whole structure. Leveraging CTI, they demonstrated the attainment of a superior loading capacity in tension compared to unidirectional composites~\cite{Hyer1987}.

Boosted by the advances in manufacturing technologies, several recent works have focused on the use of curvilinear fiber paths to attain desired mechanical properties such as e.g. higher buckling loads, lower stress concentration factors, and higher fundamental frequencies~\cite{Hyer1991,Gurdal2008,Cho2007,Cho2009,HondaS.2010Mosc,honda2013multi,Vijayachandran12020,Vijayachandran22020,mat16,Mat20}. 
Vijayachandran et al. introduced optimal fiber paths that maximize the critical buckling load of a plate under 2:1 bi-axial in-plane compressive loading by explicitly accounting for the manufacturing constraints and imperfections of an AFP machine such as e.g. tow width, tow curvatures, gaps and overlaps~\cite{Vijayachandran12020,Vijayachandran22020}. 

Along with the AFP technology, recent developments in 3D printing methods for continuous fiber-reinforced thermoplastics have enabled mold-free fabrication of composite structures. Matsuzaki et al. used fused-deposition modeling to print tensile specimens with carbon and jute fiber reinforcements following unconventional paths \cite{mat16}. Utilizing a similar technology, Sugiyama et al. designed and manufactured variable fiber volume fraction and stiffness composites (VVfSC). Fiber paths were optimized following the principal stress directions. They showed that the VVfSC plate with a hole exhibits superior specific stiffness and strength compared to conventional linear laminates \cite{Mat20}. 

Recently, Salviato and Phenisee formulated a mathematical framework for the calculation of the electric and thermal conductivities in materials featuring curvilinear transverse isotropy (CTI). They showed that the fiber paths can be designed to surpass the conductivity of any given traditional orthotropic medium made of the same materials~\cite{salviato2019}.

These examples out of many demonstrate the potential superior mechanical behavior enabled by curvilinear anisotropy compared to traditional fiber-reinforced composites.

Due to the complexities of the material systems featuring curvilinear anisotropy made via e.g. Additive Manufacturing (AM) or Automated Fiber Placement (AFP), the formulation of efficient simulation tools is still an open research topic. Isogeometric Analysis introduced by Hughes et al.~\cite{hughes2005isogeometric} is a particularly interesting candidate thanks to the following two benefits: a) better integration with Computer-Aided Design (CAD) used for additive manufactured parts and b) higher continuity between elements compared to traditional FEM. In NURBS-based Isogeometric Analysis (IGA), the overall modeling-analysis process can be significantly streamlined by employing the same Non-Uniform Rational B-Splines (NURBS) used in CAD also to perform the structural analysis \cite{nguyen2015isogeometric}. 
This avoids remeshing and increases accuracy in the geometrical description of the domain to be simulated. In addition, in NURBS it is easy to control the continuity between elements by the implementation of order elevation~\cite{hughes2005isogeometric,cottrell2009isogeometric}. This is in contrast to classical FEA which provides $C^0$ continuity between elements. Flexibility in controlling the continuity of NURBS basis functions allows to implement high-order deformation plate theories without shear correction factor such as an inverse tangent shear deformation theory demonstrated by Thai et al. \cite{Bordas3}. 
Furthermore, higher continuity in NURBS basis functions can efficiently prevent the prevalent shear locking problem as highlighted by Da Veiga et al. \cite{locking} and Valizadeh et al. \cite{Bordas2}. These advantages along with the high inter-continuity between elements are considerably desired for producing an accurate field solution in analyzing smooth curvilinear anisotropic plates.   


In this work, we first implement curvilinear anisotropy into an element stiffness matrix in the IGA framework, and then we investigate its performance by comparing it to FEA in terms of error of the stress measured in $L^2$-norm. We show that IGA has superior performance in terms of computational efficiency, time consumption, and accuracy. Then, we explore the exceptional mechanical behavior of curvilinear fiber composites in terms of in-plane stiffness and stress concentration using a plate weakened by a semi-circular notch as a case study. Finally, we identify the fiber paths minimizing the stress concentration factor and the Tsai-Wu failure index using Sequential Quadratic Programming (SQP).

\section{Mathematical background}
In the classical FEA framework, Lagrangian basis functions are utilized and defined in the physical space or in the natural space~\cite{Logan2016}. On the contrary, in the IGA framework, Non-Uniform Rational B-Spline basis functions, which integrate a modeling process (CAD) and an analysis as one package, are commonly used and defined in the parameter space~\cite{hughes2005isogeometric,cottrell2009isogeometric}. Thus, in the following sections we first provide a brief introduction on NURBS. Then, we describe the method to impose continuous fibers following curvilinear path under the isogeometric framework.

\subsection{B-splines and NURBS}
NURBS-based isogeometric analysis utilizes Non-Uniform Rational B-Spline (NURBS) basis functions defined by a knot vector:
\begin{equation}\label{eq:Xi_knot}
    \boldsymbol{\Xi}=\begin{bmatrix}
    \xi_{1}, \xi_{2}, \xi_{3}, \cdots,\xi_{n+p+1}\\
    \end{bmatrix}, \qquad \text{with} \quad \xi_{i} \in \mathbb{R}.
\end{equation}
The knot vector is a non-decreasing set of coordinates in the parameter space, where $i = 1,\cdots,n+p+1$, and $n$ is the number of basis functions of the order $p$ used to construct the B-spline curve, which were first proposed by Gordon and Riesenfeld~\cite{gord74,Riesenfeld1973} for computer-aided graphic design. Given the knot vector in the parameter space defined above, the $i^{th}$ B-spline basis function of order $p$, denoted by $N_{i,p}(\xi)$ can be computed following Cox-de Boor recursion formula~\cite{Cox72,DEBoor1972,de1978practical} as shown in Eq. (\ref{b-spline_p0}) and (\ref{b-spline}): \\
For $p = 0:$
\begin{equation}\label{b-spline_p0}
N_{i,0}(\xi) = \left\{
\begin{array}{rl}
1 & \text{if } \xi_{i} \leq \xi < \xi_{i+1},\\
0 & \text{otherwise,}
\end{array} \right.
\end{equation}
for $p = 1,2,3,\cdots,$
\begin{equation}\label{b-spline}
N_{i,p}(\xi) = \frac{\xi-\xi_{i}}{\xi_{i+p}-\xi_{i}}N_{i,p-1}(\xi)
+\frac{\xi_{i+p+1}-\xi}{\xi_{i+p+1}-\xi_{i+1}}N_{i+1,p-1}(\xi).
\end{equation}
By taking a linear combination of the B-spline basis $N_{i,p}(\xi)$ and the corresponding control points $P_i \in \mathbb{R}$, the B-spline curve is described as
\begin{equation}\label{curve}
\mathbf{C}(\xi) = \sum^n_{i = 1}N_{i,p}(\xi)P_{i}, \quad 0 \leq \xi_{i} \leq 1.
\end{equation}
Similarly, assigning a control net $P_{i,j}$, where $i = 1,...,n$, $j = 1,...,m$, and B-spline functions in $\xi$- and $\eta$-directions $N_{i,p}(\xi)$ and $M_{j,q}(\eta)$ with the order of $p$ and $q$ constructed by knot vectors $\boldsymbol{\Xi} = [\xi_1,...,\xi_{n+p+1}]$, and $\boldsymbol{H} = [\eta_1,...,\eta_{m+q+1}]$, respectively, the B-spline surface is also given by
\begin{equation}\label{surface}
\mathbf{S}(\xi,\eta) = \sum^n_{i = 1}\sum^m_{j = 1}
N_{i,p}(\xi)M_{j,q}(\eta)P_{i,j}, \quad 0 \leq \xi_{i} \leq 1 \quad \text{and} \quad 0 \leq \eta_{j} \leq 1
\end{equation}

Making a rational function divided by weight functions enables exact representations for a wide range of complex geometry, especially in conic sections (circles, ellipses, etc.), which are beyond an array of piece-wise polynomials. This rational function is referred to as Non-Uniform Rational B-Spline function or NURBS (See the references e.g.~\cite{Vers75,tiller83,Piegl1991}). Given the knot vector $\boldsymbol{\Xi}$ defined above, the $p^{th}$-order NURBS function on the knot span $\xi \in [\xi_i,\xi_{i+1})$ is given by
\begin{equation}\label{nurbs}
R^p_i(\xi) = \frac{N_{i,p}(\xi)\omega_i}{\sum^n_{\hat{i}=1}N_{\hat{i},p}\omega_{\hat{i}}}
\end{equation}
where $\omega_i$ is the $i^{th}$ weight and $N_{i,p}(\xi)$ is the $i^{th}$ B-spline function with order $p$.
Then, the NURBS curve and surface based on the knot vectors $\boldsymbol{\Xi}$ and $\boldsymbol{H}$ can be generated using the NURBS basis functions as shown in Eq. (\ref{nurbs}), respectively:
\begin{equation}\label{nurbs_curve}
\mathbf{C}(\xi) = \sum^n_{i = 1}R_i^p(\xi)P_i,
\end{equation}
and
\begin{equation}\label{nurbs_surface}
\mathbf{S}(\xi,\eta) = 
\sum^n_{i= 1}\sum^m_{j=1}R_{i,j}^{p,q}(\xi)P_{i,j}.
\end{equation}
where
\begin{equation}\label{bi-variable_nurbs}
R^{p,q}_{i,j}(\xi,\eta) = \frac{N_{i,p}(\xi)M_{j,q}(\eta)\omega_{i,j}}
{\sum^n_{\hat{i}=1}\sum^m_{\hat{j}=1}N_{\hat{i},p}M_{\hat{j},q}(\eta)\omega_{\hat{i},\hat{j}}},
\end{equation}
$R^{p,q}_{i,j}(\xi,\eta)$ is a bi-variable NURBS basis function and $\omega_{i,j}$ is a weight for the corresponding control net $P_{i,j}$.

\subsection{Derivative of B-Splines and NURBS}
Defining the $p^{th}$-order B-spline function $N_{i,p}(\xi)$ by the knot vector $\boldsymbol{\Xi} = [\xi_1,...,\xi_{n+p+1}]$, the derivative $N'_{i,p}(\xi)$, is calculated recursively as:
\begin{equation}\label{deriv_b-splime}
N'_{i,j}(\xi) =  \frac{p}{{\xi_{i+p}-\xi_{i}}}N_{i,p-1}(\xi)
-\frac{p}{\xi_{i+p+1}-\xi_{i+1}}N_{i+1,p-1}(\xi).
\end{equation}
This can be proven by a mathematical induction~\cite{piegl1997nurbs}.

Starting from the derivative of B-Splines, the derivative of NURBS functions is simply derived using the quotient rule to Eq. (\ref{nurbs}) such that:
\begin{equation}\label{deriv_nurbs}
\frac{d}{d\xi}R^p_i(\xi) = \omega_i\frac{W(\xi)N'_{i,p}(\xi)-W'(\xi)N_{i,p}(\xi)}{[W(\xi)]^2},
\end{equation}
where
\begin{equation}\label{W}
W'(\xi) = \sum^n_{\hat{i}=1}N'_{\hat{i},p}(\xi)\omega_{\hat{i}}.
\end{equation}

Further properties and techniques for B-Splines and NURBS can be found in~\cite{de1972calculating,Mortenson1985,Farin1993,Rogers1986} and~\cite{piegl1997nurbs,Vers75,tiller83,Piegl1991}, respectively. For detailed theoretical framework and the implementation in IGA, the reader is referred to Hughes et al.~\cite{hughes2005isogeometric,cottrell2009isogeometric}.

\subsection{Curvilinear fiber description for simple optimization studies}
Typically, the description of curvilinear fiber paths is performed assigning a constant tangent vector in each element and considering each element orientation as a variable of the parameter space ~\cite{Hyer1987,Hyer1991,Cho2007,Cho2009}. While this approach can be very accurate, it also leads to a significant computational cost due to the large number of design variables (the number of design variables is equal to the number of elements). Since the main goal of the optimization study proposed here is to showcase the use of IGA rather than identifying the perfect fiber configuration, we opted for the simplified approach proposed in \cite{HondaS.2010Mosc,honda2013multi}. Hence, we utilize the concept of level-set function~\cite{HondaS.2010Mosc,honda2013multi} to present a subset of possible fiber configurations in the chosen domain, which also ensures the following important characteristics in terms of manufacturability: 1) continuity of fiber paths, and 2) absence of intersections between fibers. Therefore, in the present work, we describe the curvilinear fiber paths by using a $5^{th}$-order complete polynomial as a level-set function:
\begin{equation}\label{level}
	\begin{split}
	f(x,y) = C_{00} + &C_{10}x + C_{01}y + C_{20}x^2 + C_{11}xy + C_{02}y^2 \\ 
	& + C_{30}x^3 + C_{21}x^2y + C_{12}xy^2 + C_{03}y^3 \\
	& + C_{40}x^4 + C_{31}x^3y + C_{22}x^2y^2 + C_{13}xy^3 + C_{04}y^4 \\
	& + C_{50}x^5 + C_{41}x^4y + C_{32}x^3y^2 + C_{23}x^2y^3 + C_{14}xy^4 + C_{05}y^5
	\end{split}	
\end{equation}
where $C_{ij}$, with $i,j = 0,...,5$, are design variables of the $5^{th}$-order complete polynomial function. The fiber orientations $\theta$ are also calculated by partially differentiating the level set function of Eq. (\ref{level}) with respect to $x$ and $y$ as:
\begin{equation}\label{angle}
	\theta(x,y) = \tan^{-1}\Bigg(-\frac{\partial f/\partial x}{\partial f/\partial y}\Bigg)
\end{equation}

It is important to restate here that the optimization study only serves the purpose of demonstrating the IGA framework. The level-set functions can only represent a subset of the fiber path solutions and cannot be exhaustive. Ongoing work by the authors is focusing on robust implementation of a comprehensive optimization framework. However, its description goes beyond the scope of the present work.

\subsection{Element implementation of the elastic behavior for CTI materials}
The constitutive relation of CTI media depends on the local fiber orientation. As a consequence, the element stiffness matrix must be evaluated at each Gauss point. The constitutive relations for a linear elastic problem can be expressed as:
\begin{equation}\label{system}
    \{F\} = [K]\{d\}
\end{equation}
where $\{F\}$, $\{d\}$, and $[K]$ are the loading vector, the displacement vector, and the global stiffness matrix, respectively. Then, the element stiffness matrix in the physical domain $\Omega^{(e)}$ can be written in the form:
\begin{equation}\label{elem_stiffness}
    [K]^{(e)} = \int_{\Omega^{(e)}}[B]^T[\overline{C}]^{(\theta)}[B] \,d\Omega^{(e)},
\end{equation}
where $[B]$ is the strain-displacement matrix generated from the NURBS basis functions, Eq. (\ref{bi-variable_nurbs}), and
\begin{equation}\label{modified_stiffness}
     [\overline{C}]^{(\theta)} = [T_{\sigma}^{-1}(\theta)][C][T_{\epsilon}(\theta)],
\end{equation}
where $[T_{\sigma}^{-1}(\theta)]$ and $[T_{\epsilon}^{-1}(\theta)]$ are transformation matrices for the stress and strain, respectively, and $[C]$ is the in-plane stiffness coefficients matrix depending on three engineering moduli ($E_1$, $E_2$, and $G_{12}$), and the Poisson's ratio ($\nu_{12}$)~\cite{Gibson1994,kol03}. Then, the element stiffness is calculated by Gauss quadrature such that:
\begin{equation}\label{elem_stiffness_integral}
\left[K\right]^{(e)}=\sum_{m=1}^{N_p}w_{\tilde{\xi}_m}w_{\tilde{\eta}_m}[B_m]^T[\overline{C_m}]^{(\theta_m)}[B_m]J_m,
\end{equation}
where $N_p$ is the number of integration points, $w_{\tilde{\xi}_m}$ and $w_{\tilde{\eta}_m}$ are the corresponding weighs for the each direction, and $J_m$ is the determinant of the Jacobian which maps from physical space to parent space. Detail derivation and calculation of the element stiffness in the isogeometric framework can be found in~\cite{nguyen2015isogeometric}.

\begin{figure}
    \includegraphics[scale=.75]{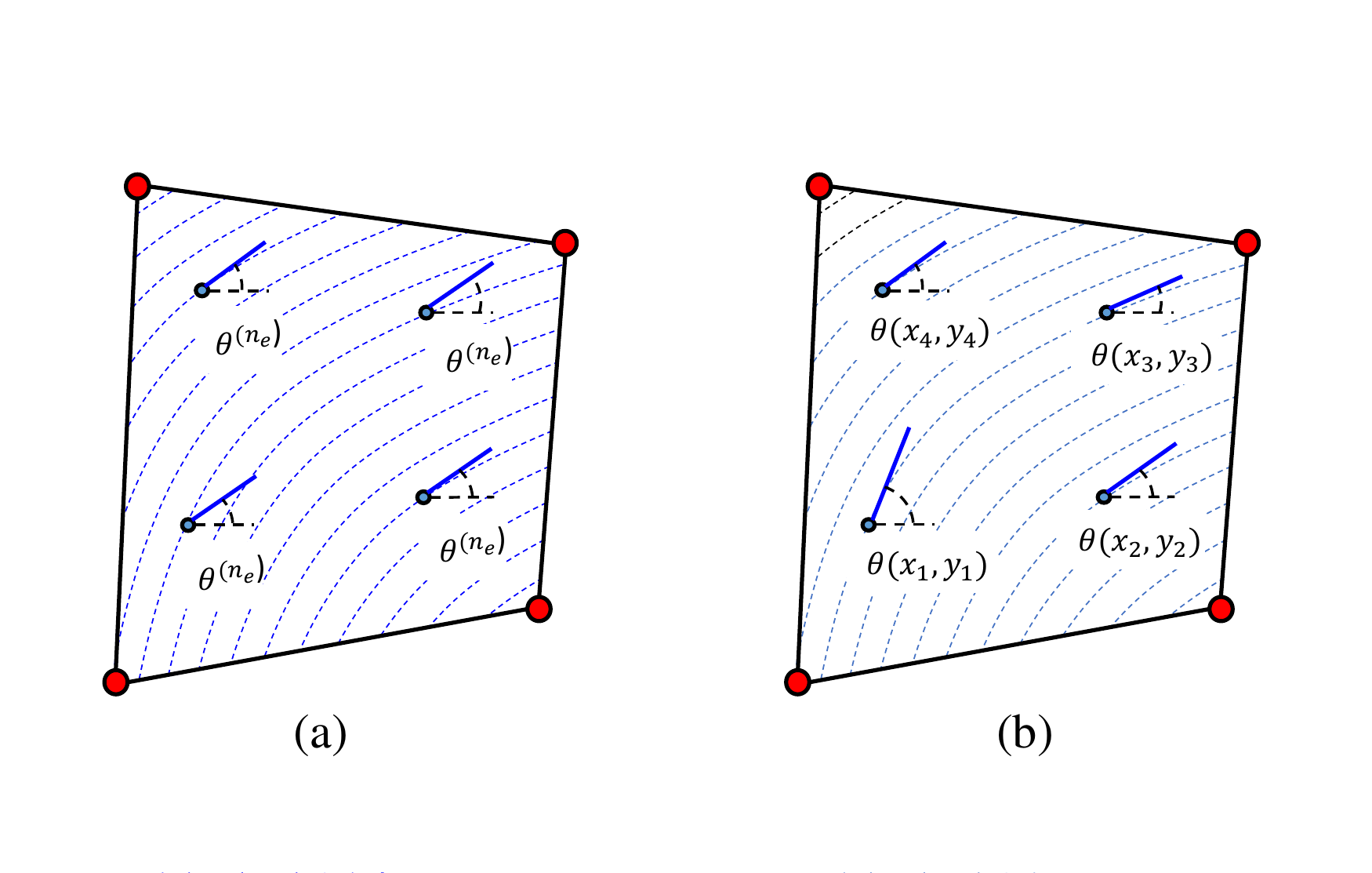}
    \centering
    \caption[Schematic representation of 4-node, isoparametric, quadrilateral element]{Schematic representation of 4-node, isoparametric, quadrilateral element with material orientations assigned to each integration point. As can be noted, fiber orientations are assigned as constant in previous works: (a). In this present work, we let the fiber orientation depend on the location of integration points during the computation of element stiffness matrix: (b).}
    \label{integration}
\end{figure}

As Fig. \ref{integration} shows, the analysis of CTI media in previous works~\cite{Hyer1987,Hyer1991,HondaS.2010Mosc,honda2013multi,Cho2007,Cho2009} was performed using a uniform constitutive relation for the integration points in each element. In other words, $\theta = \theta^{(n_e)}$, with $n_e$ is an element number, i.e., $[K]^{(e)} = [K^{(e)}\{\theta^{(n_e)}\}]^{(e)}$. In contrast, in the present work we let the constitutive relation depend on the location of integration points within the element during the computation of element stiffness matrix such as $\theta = \theta(x,y)$, where $x$ and $y$ are the coordinates of the integration points, i.e., $[K]^{(e)} = [K\{\theta(x,y)\}]^{(e)}$~\cite{Suzuki2019}. This results in accounting for the effects of the gradient in fiber curvature in calculating the element stiffness.


\section{Optimization method}
\subsection{Sequential quadratic programming method}
For the design optimization study in this paper, we adopted Sequential Quadratic Programming (SQP) assuming that the object function is twice differentiable. The SQP is one of the most powerful optimization methods designed for Constrained Non-Linear Programming (NLP) problems~\cite{Arora2012,Nocedal2006}. This enables to achieve an attractive rate of convergence using $2^{nd}$-order derivatives of Lagrange function.

In this paper, the SQP follows the Constrained Steepest-Descent (CSD) method via two major procedures such that 1) Quadratic Programming (QP) subproblem and 2) Quasi-Newton method. Let define an NLP problem of the form:

Minimize
\begin{equation}\label{object_1}
    f(\boldsymbol{x}),
\end{equation}

subject to
\begin{equation}\label{bcs_1}
\begin{split}
    & \qquad h_i(\boldsymbol{x}) = 0; \quad i = 1 \quad \text{to} \quad p, \\
    & \qquad g_i(\boldsymbol{x}) \leq 0; \quad i = 1 \quad \text{to} \quad m.
\end{split}
\end{equation}
In the SQP method, QP subproblem is solved for a search direction $\boldsymbol{d}$ in a CSD framework. Thus, Karush-Kuhn-Tucker necessary conditions (KKT conditions)~\cite{karush39,Kuhn1951} can be applied utilizing the Newton-Raphson method, and a general constrained optimization problem $\bar{f}$ is derived as:

Minimize
\begin{equation}\label{object_2}
    \bar{f} = \boldsymbol{c}^T\boldsymbol{d} + \frac{1}{2}\boldsymbol{d}^T\boldsymbol{H}\boldsymbol{d}
\end{equation}

subject to the constraints:
\begin{equation}\label{bcs_2}
\begin{split}
    & \boldsymbol{n}^{(i)^{T}}\boldsymbol{d} = -h_i(\boldsymbol{x}^{(k)}); \qquad i = 1 \quad \text{to} \quad p, \\
    & \boldsymbol{a}^{(i)^{T}}\boldsymbol{d} \leq -g_i(\boldsymbol{x}^{(k)}); \qquad i = 1 \quad \text{to} \quad m,
\end{split}
\end{equation}
where $\boldsymbol{H}$ is the Hessian matrix of Lagrange function $\nabla^2{L}$, $\boldsymbol{n}^{(i)}$ and $\boldsymbol{a}^{(i)}$ are the gradients of the functions $h_i$ and $g_i$, respectively, $\boldsymbol{c}$ is the gradient of the objective function of Eq. (\ref{object_1}), and $\boldsymbol{x}^{(k)}$ represents the current iteration point. Thus, the search direction $\boldsymbol{d}$ is computed by the gradient conditions of the KKT necessary conditions.

Once the QP subproblem has been derived, the Hessian matrix $\boldsymbol{H}$ must be approximated for each iteration of the CSD framework. The Hessian matrix $\boldsymbol{H}$ is approximated using Broyden-Fletcher-Goldfarb-Shanno (BFGS) method~\cite{Broyden1970,Fletcher1970,Goldfarb1970,Shanno1970}. Define the gradient difference vector $\boldsymbol{y}^{(k)}$ of the Lagrange function at two iteration points and the design change vector $\boldsymbol{s}^{(k)}$ with the Lagrange multipliers $\boldsymbol{v}^{(k)}$ and $\boldsymbol{u}^{(k)}$ at the current point such that
\begin{equation}\label{grad_diff_vec}
\begin{split}
    & \boldsymbol{y}^{(k)} = \nabla{L(\boldsymbol{x}^{(k+1)},\boldsymbol{v}^{(k)},\boldsymbol{u}^{(k)})} - \nabla{L(\boldsymbol{x}^{(k)},\boldsymbol{v}^{(k)},\boldsymbol{u}^{(k)})}, \\
    & \boldsymbol{s}^{(k)} = \alpha_k\boldsymbol{d}^{(k)}; \quad \alpha_k = \text{step size}.
\end{split}
\end{equation}
and set the vector $\boldsymbol{r}^{(k)}$ with the scalar quantity $\theta$ as:
\begin{equation}\label{r_vec}
    \boldsymbol{r}^{(k)} = \theta^{(k)}\boldsymbol{y}^{(k)}+(1-\theta^{(k)})\boldsymbol{H}^{(k)}\boldsymbol{s}^{(k)},
\end{equation}
\begin{equation}\label{theta_quantity}
    \theta^{(k)} = \left\{
    \begin{array}{ll}
    1; \quad \text{if } \boldsymbol{s}^{(k)^T}\boldsymbol{y}^{(k)} \geq 0.2\boldsymbol{s}^{(k)^T}\boldsymbol{H}^{(k)}\boldsymbol{s}^{(k)},\\
    (0.8\boldsymbol{s}^{(k)^T}\boldsymbol{H}^{(k)}\boldsymbol{s}^{(k)})/(\boldsymbol{s}^{(k)^T}\boldsymbol{H}^{(k)}\boldsymbol{s}^{(k)}-\boldsymbol{s}^{(k)^T}\boldsymbol{y}^{(k)}); \quad \text{otherwise.}
    \end{array} \right.
\end{equation}
Then, the Hessian matrix can be updated as:
\begin{equation}\label{opt1}
    \boldsymbol{H}^{(k+1)} = \boldsymbol{H}^{(k)} + \frac{\boldsymbol{r}^{(k)}\boldsymbol{r}^{(k)^T}}{\boldsymbol{r}^{(k)^T}\boldsymbol{s}^{(k)}}-\frac{\boldsymbol{H}^{(k)}\boldsymbol{s}^{(k)}\boldsymbol{s}^{(k)^T}\boldsymbol{H}^{(k)^T}}{\boldsymbol{s}^{(k)^T}\boldsymbol{H}^{(k)}\boldsymbol{s}^{(k)}}
\end{equation}

Finally, the solutions of Eq. (\ref{object_1}) can be obtained by the use of above two methods 1) SQP subproblem and 2) Quasi-Newton method in the CSD framework. Further explanations including schematic algorithms on these methods can be found in~\cite{Nocedal2006,Arora2012}.

\subsection{Optimal fiber design problem}
To demonstrate the IGA framework, the goal of our optimization study is obtaining optimal curvilinear or linear fiber paths with a minimum stress concentration or Tsai-Wu failure index. Thus, assuming that our objective functions $K_t$ for a minimum stress concentration factor and a minimum Tsai-Wu failure index $\Phi$ are both twice differentiable nonlinear functions, the optimal design problems can be defined as

Minimize
\begin{equation}\label{opt2}
	K_t(C_{ij}),
\end{equation}
\begin{equation}\label{opt3}
	\Phi(C_{ij}),
\end{equation}
where $C_{ij} \in \mathbb{R}$ are the coefficients of the level-set function defined in Eq. (\ref{level}). Thus, the objective functions are subject to the fiber orientations in Eq. (\ref{angle}). In this study, the SQP method was implemented in the MATLAB environment utilizing built-in function called \textit{fmincon} with an option of \textit{sqp} algorithm~\cite{Mathworks2020}.

\section{Results and discussion}
\begin{figure}[!htbp]
	\includegraphics[scale=0.7]{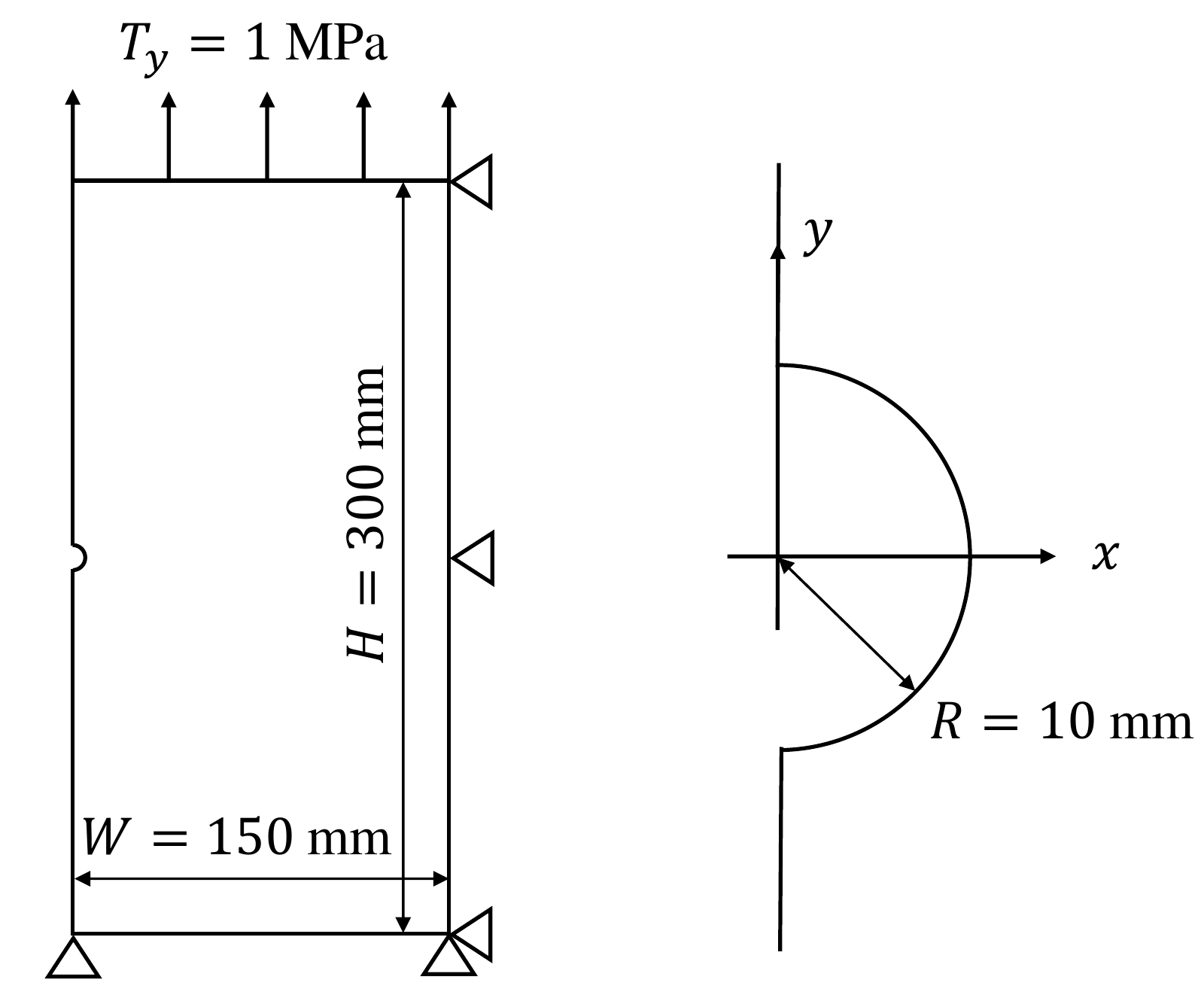}
	\centering
	\caption[Elastic semi-circular hole notched plate with a radius of $10$ mm and an unit thickness]{Elastic plate featuring curvilinear transverse isotropy weakened by a $10$-mm semi-circular hole notch.}
	\label{plate}
\end{figure}

\begin{figure}[!htbp]
    \centering
	\includegraphics[scale=0.75]{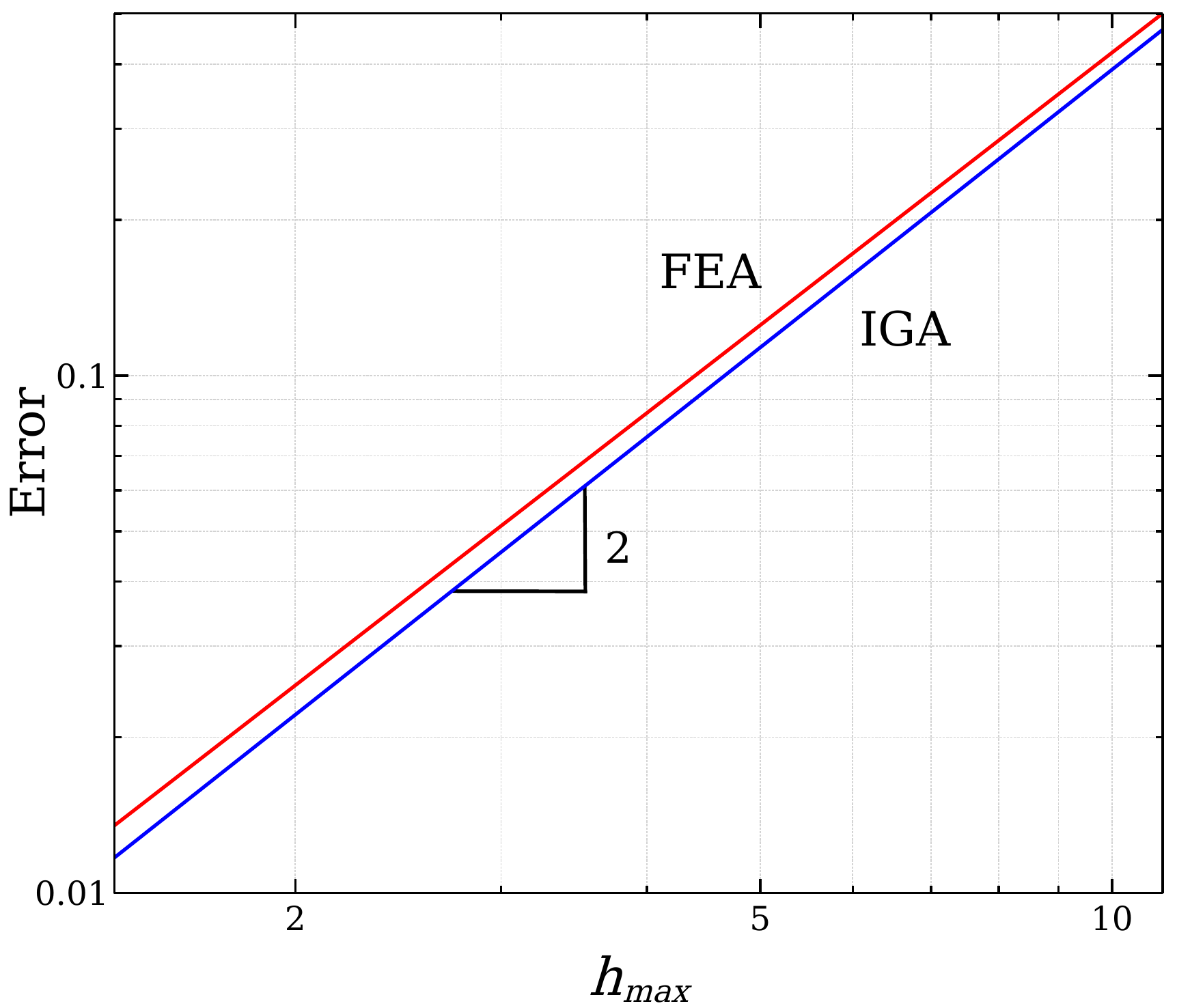}
	\caption[Error measured in $L^2$-norm of the stress against the maximum diagonal length found in the mesh ($h_{max}$)]{Error measured in $L^2$-norm of the stress against the maximum diagonal length found in the mesh $h_{max}$.}
	\label{convergence_hmax}
\end{figure}

In this section, Isogeometric Analysis (IGA) is implemented on fiber reinforced, semi-circular notched plates modeled by Non-Uniform Rational B-Splines (NURBS) functions. The IGA solver is adapted from~\cite{nguyen2015isogeometric} to simulate orthotropic media in MATLAB environment~\cite{Suzuki2019}. Finite Element Analysis (FEA) is also implemented on the semi-circular notched plates with several fiber configurations. The FEA solver is generated in MATLAB environment and validated by analytical solution of an infinite plate with a circular hole. Both IGA and FEA were also verified by checking the rates of convergence on the simulation of the elastic, isotropic plate described in Fig. \ref{plate}. The stress errors computed in $L^2$-norm were plotted as a function of the maximum diagonal length found in the mesh in double-logarithmic scale. As shown in Fig. \ref{convergence_hmax}, the slopes are both about 2 for the quadratic element, matching perfectly the theoretical values~\cite{cottrell2009isogeometric}. The convergence of IGA is discussed highlighting a superior numerical performance compared to classical FEA. Then, we present compelling results on the mechanical behavior of composite materials featuring CTI.

\subsection{Problem setup}

Stress analysis on an elastic plate weakened by a semi-circular hole notch is conducted using both NURBS-based IGA and classical FEA assuming a linear elastic problem illustrated in Fig. \ref{plate}. The semi-circular notched plate is under a unit constant tension of $T_y = 1$ MPa along the top edge and assumed as a plane-stress problem with a unit thickness. The plate is constrained along the bottom and the right-hand side edges in the $y$- and the $x$-directions, respectively. The elastic plate is modeled using quadratic NURBS basis functions in the IGA simulations with the two knot vectors $\boldsymbol{\Xi} = [0,0,0,0.25,0.5,0.5,0.75,1,1,1]$ and $\boldsymbol{H} = [0,0,0,1,1,1]$, and mesh refinement is implemented using knot insertion~\cite{hughes2005isogeometric,cottrell2009isogeometric}. In conventional FEA, simulations are conducted applying quadrilateral 9-node element (Q9) and, the exact same node coordinates used in IGA are employed for the sake of fair comparison in terms of convergence. The semi-circular notched plate with several fiber configurations is examined by the following properties:
\begin{enumerate}
    \item Stress concentration factor $K_t$:
    \begin{equation}\label{Kt}
        K_t = \sigma_{y}^{max}/T_y, \quad \sigma_{y}^{max} \text{: Maximum longitudinal stress found in the body,}
    \end{equation}
	\item Average in-plane stiffness along the top edge $\bar{K}$:
	\begin{equation}\label{stiffness}
		\bar{K} = \frac{P_{tot}}{\bar{v}}, \quad \text{with} \quad
		P_{tot} = \int_{0}^{L}\sigma_{yy} d\,x \quad \text{and} \quad \bar{v} = \frac{1}{L} \int_{0}^{L} v d\,x,
	\end{equation}
	where $\sigma_{yy}$ and $v$ are the longitudinal stress and displacement along the top edge, respectively,  and $L$ is the length of the top edge.
	\item Tsai-Wu failure index $\Phi$ assuming Transverse Isotropy under plane-stress condition~\cite{Tsai1971}:
	\begin{equation}\label{tsai_wu}
	 	\Phi = F_1\sigma_{11} + F_2\sigma_{22} + F_{11}\sigma_{11}^2 + 	F_{22}\sigma_{22}^2 + F_{66}\tau_{12}^2 + 2F_{12}\sigma_{11}\sigma_{22},
	\end{equation}
	where
	\begin{equation}\label{tsai_wu_1}
		 \begin{split}
	 	F_1    &= \frac{1}{\sigma_{1t}^f} - \frac{1}{\sigma_{1c}^f}, \qquad F_2 = \frac{1}{\sigma_{2t}^f} - \frac{1}{\sigma_{2c}^f}, \qquad F_{66} = \frac{1}{\tau_{12}^f}, \\
	 	F_{11} &= \frac{1}{\sigma_{1t}^f \sigma_{1c}^f}, \qquad F_{22} = \frac{1}{\sigma_{2t}^f \sigma_{2c}^f}, \qquad F_{12} = -\frac{1}{2}\sqrt{F_{11}F_{22}}. \\
	 	\end{split}
	\end{equation}
\end{enumerate}
The choice of this criterion is motivated by its effectiveness and accuracy in capturing the failure condition of composite laminates featuring smooth or slightly notched surfaces. However, for more complex structural configurations and loading conditions, the emergence of large Fracture Process Zones (FPZs) can lead to significant size effects. This is typically the case of structures featuring open and filled holes, sharp notches and other stress raisers. In such a case, a quasibrittle fracture mechanics framework should be preferred over a stress-based failure criterion \cite{bazant, Salviato2016,salviato1,salviato2,okabe2020}. All the required material properties such as longitudinal and transverse elastic moduli $E_1$ and $E_2$, shear modulus $G_{12}$, Poisson's ratio $\nu_{12}$, and the ply strengths: $\sigma_{1t}^f$, $\sigma_{1c}^f$, which are tensile and compressive strengths in the fiber direction, $\sigma_{2t}^f$, $\sigma_{2c}^f$, which are tensile and compressive strengths perpendicular to the fiber directions, and $\tau_{12}^f$, that is shear strength are respectively assigned and tabulated in Table \ref{eng_constants}.

\subsection{Fiber configurations}
In the present study, four different fiber configurations as described in Fig. \ref{paths} were investigated: a) conventional longitudinal straight fiber paths, b) conventional transverse straight fiber paths, c) concentric fiber paths following a semi-circular notch, and d) curvilinear fibers following the holomorphic path defined by the conformal mapping presented in~\cite{salviato2019}. 
The four fiber orientations measured from the global $x$-axis are defined as a function of $(x,y)$ coordinate such that:
\setlist{nolistsep}
\begin{enumerate}[label=(\alph*)]
    \item Conventional longitudinal straight fibers:
    $\theta(x,y) = \pi/2,$
    \item Conventional transverse straight fibers:
    $\theta(x,y) = 0,$
    \item Concentric fibers following the semi-circular hole:
    \begin{equation*}
        \theta(x,y) = \frac{\pi}{2} + \cos^{-1}{\frac{x}{a}}, \quad a = \sqrt{x^2+y^2},
    \end{equation*}
    \item Curvilinear fibers following the holomorphic function presented in ~\cite{salviato2019}:
    \begin{equation*}
        \theta(x,y) = \tan^{-1}\Bigg[\frac{1+\frac{R^2}{x^2+y^2}\cos\big(2\tan\frac{y}{x}\big)}{-\frac{R^2}{x^2+y^2}\sin\big(2\tan\frac{y}{x}\big)}\Bigg].
    \end{equation*}
\end{enumerate}

\subsection{Performance of Isogeometric Analysis}
\begin{figure}[!htbp]
    \centering
	\includegraphics[scale=0.76]{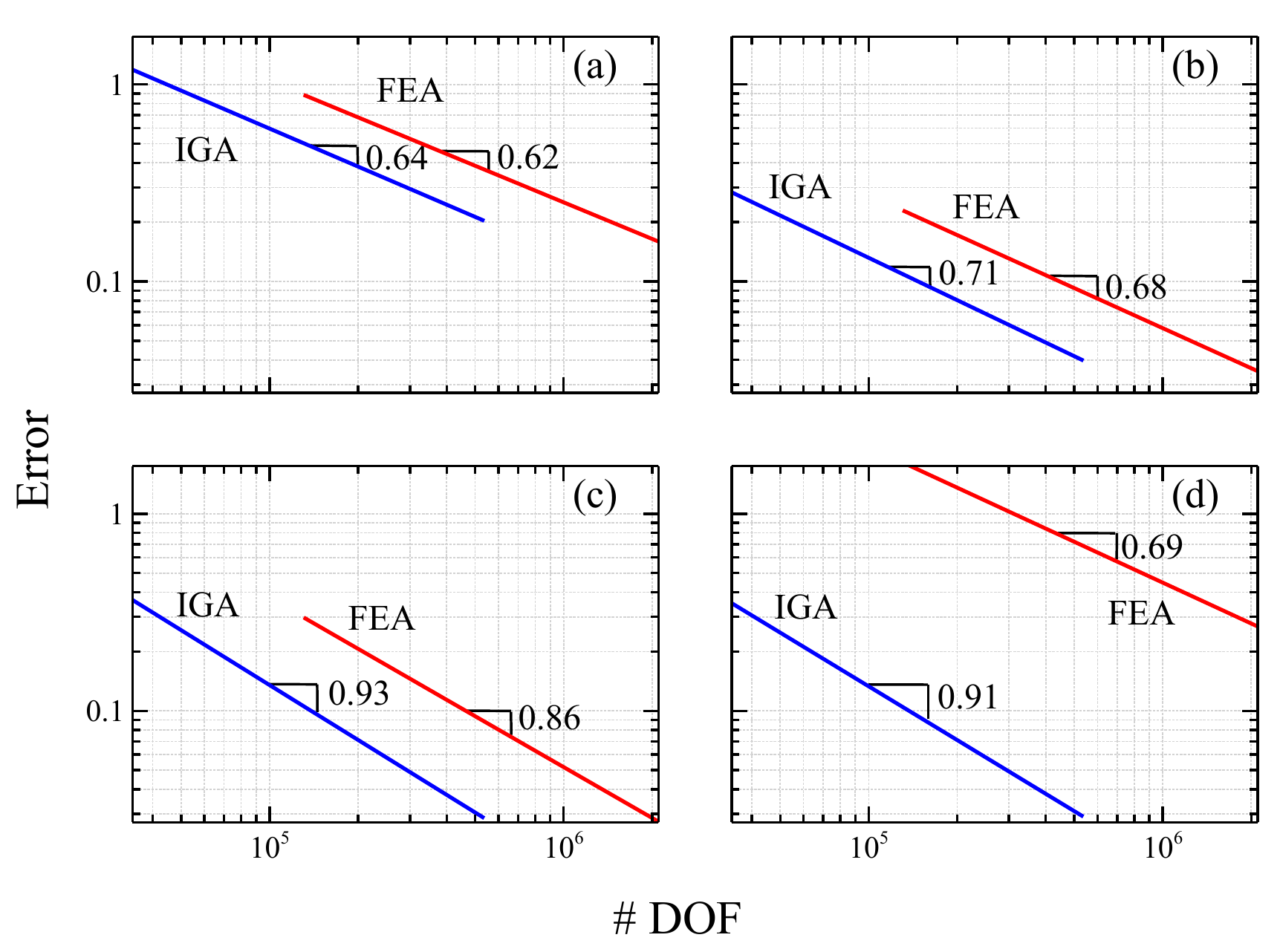}
	\caption[Error measured in $L^2$-norm of the stress against the number of degrees-of-freedom ($\#$DOF)]{Error measured in $L^2$-norm of the stress against the number of degrees-of-freedom ($\#$ DOF): (a) Longitudinal straight fiber paths, (b) Transverse straight fiber paths, (c) Concentric fiber paths following the semi-circular notch, (d) Curvilinear fiber paths following the semi-circular notch.}
	\label{convergence}
\end{figure}

Figure \ref{convergence} shows the error computed in $L^2$-norm of stresses for each of the fiber configurations presented in Fig. \ref{paths}. The error for each component is defined by the following:
\begin{equation}\label{L2_error}
    \boldsymbol{\epsilon} = \sqrt{\oint_{V}\boldsymbol{\epsilon}_{\sigma}\boldsymbol{\epsilon}_{\sigma} \,dV},
\end{equation}
where $V$ is the domain of the whole material, and the $\boldsymbol{\epsilon}_{\sigma}$ is the vector of the difference between the sufficiently converged stress field (change in the $K_t$ less than $0.5$\%) and the current stress field.
Then, the error $\epsilon$, the root-mean-value of the Eq. (\ref{L2_error}) is taken as:
\begin{equation}\label{rms}
    \epsilon = \sqrt{\frac{\boldsymbol{\epsilon}\cdot\boldsymbol{\epsilon}}{n}}
\end{equation}
where $n$ is the number of components for the stress field and ``$\cdot$" represents the dot product here.

\begin{figure}[!htbp]
    \centering
	\includegraphics[scale=0.55]{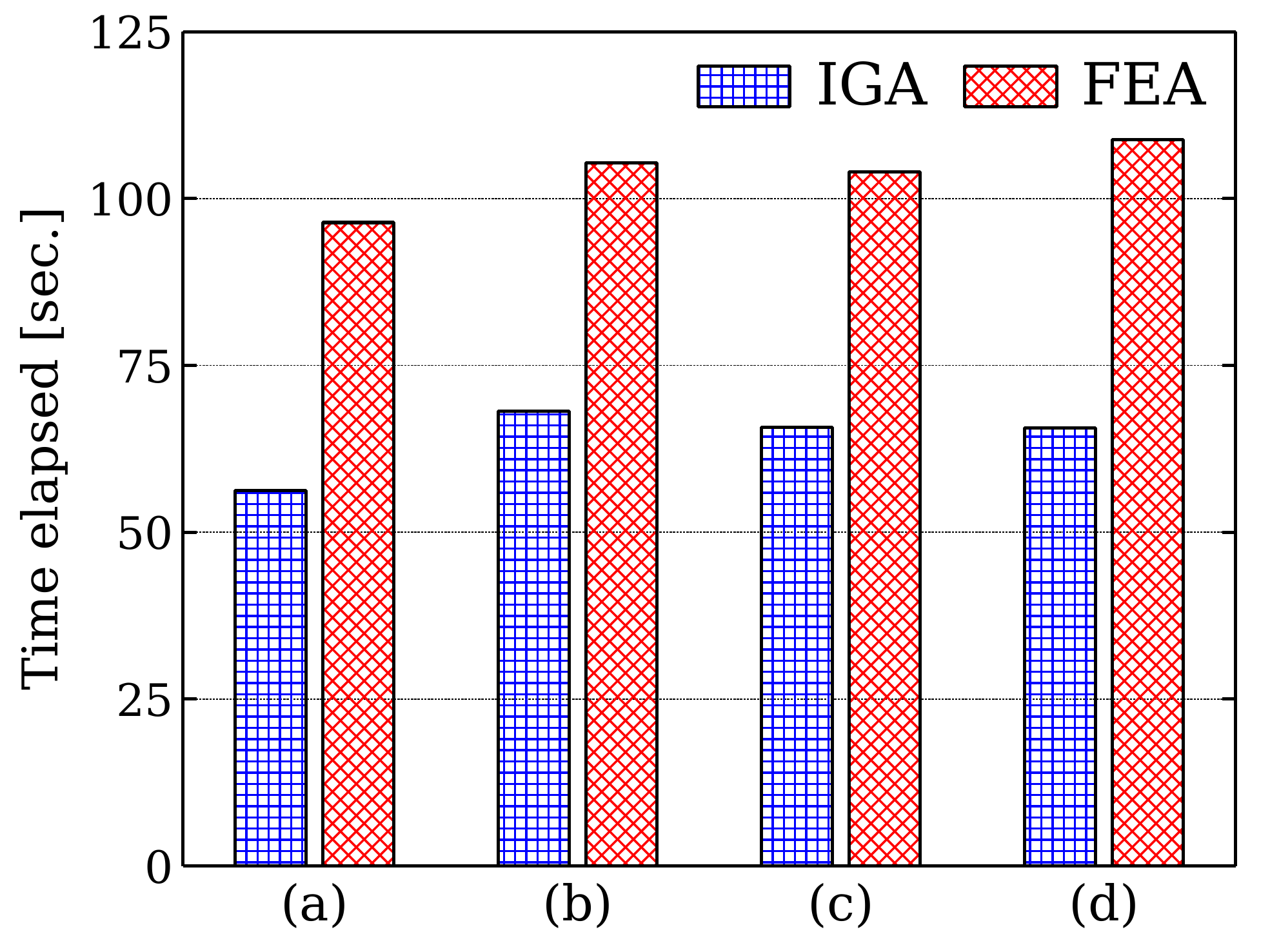}
	\caption[Running time in seconds elapsed to obtain convergence for each fiber configurations]{Running time in seconds elapsed to obtain convergence for each fiber configuration: (a) Longitudinal straight fiber paths, (b) Transverse straight fiber paths, (c) Concentric fiber paths following the semi-circular notch, (d) Curvilinear fiber paths following the semi-circular notch.}
	\label{run_time}
\end{figure}

\noindent The errors are plotted in double-logarithmic scale as a function of the number of degrees-of-freedom for both IGA and FEA. By utilizing the same matrix inversion scheme in both IGA and FEA, the running time is measured at the very last point of each convergence curve on our computer with an AMD Ryzen 5 2600 processor running at 3.4 MHz using 32 GB of RAM, and the percentage reduction is defined as:
\begin{equation}
    \Delta t\% = \frac{|t_{IGA}-t_{FEA}|}{t_{FEA}}*100
\end{equation}
where $t$ represents the running time elapsed each numerical method, and the numerator is given as an absolute value (See Fig. \ref{run_time}). The running time of IGA is approximately $35$-$40\%$ lower than conventional FEA. This means that not only IGA can reduce the time from Computer-Aided Design (CAD) to analysis, it also enables a significant reduction of the running time of an analysis. This result becomes more significant when the size and complexity of the analysis increases.

In addition, we also focused on the continuity of the basis functions for both IGA and FEA. In the present study, quadratic Lagrange polynomials are adopted as the basis function in the FEA framework, which leads to $C^0$-continuity between elements all the time. On the contrary, the basis function of NURBS-based IGA is decided by the quadratic NURBS basis function with $C^1$-continuity everywhere except along the $x$-axis due to the multiplicity of the knot values at $\xi = 0.5$. The enhanced smoothness of the interpolating functions element by element results in capturing the field variables between elements in better numerical performance for a given number of degrees-of-freedom compared to classical FEA. Moreover, as can be seen in Fig. \ref{convergence}d, the difference between the errors for IGA and FEA is way greater than the ones of the conventional fiber configurations (Fig. \ref{convergence}a and \ref{convergence}b), which indicates that IGA especially plays a significant role in the CTI specimens.

In summary, IGA provides: 
\setlist{nolistsep}
\begin{enumerate}[noitemsep]
\item time reduction
\item computational efficiency
\item more accurate description of the field variables for a given number of degrees-of-freedom, compared to classical FEA
\end{enumerate}

These features make IGA a valuable tool for the simulation of structural and electrostatics problems in media featuring curvilinear anisotropy systems~\cite{salviato2019,Hyer1987,Hyer1991,Gurdal2008,Cho2007,Cho2009,HondaS.2010Mosc,honda2013multi,Vijayachandran12020,Vijayachandran22020}.

\subsection{Investigation of the effects of curvilinear anisotropy on the elastostatic behavior}
As can be seen in Table \ref{results_1}, generally, conventional longitudinal fiber reinforced composites have high-level of in-plane stiffness with significantly high stress concentration around notches and other stress raisers. In contrast, reinforced composites featuring transverse straigth fibers have lower stress concentration and in-plane stiffness. Leveraging the IGA theoretical framework, here we explore how the different curvilinear fiber paths described in Fig. \ref{paths} affect stress concentration and stiffness. All the stress concentration factors and Tsai-Wu failure indexes obtained from simulations are tabulated in Table \ref{results_1}.

\subsubsection{Concentric fibers following semi-circular paths}
\begin{figure}[!htbp]
	\includegraphics[scale=0.35]{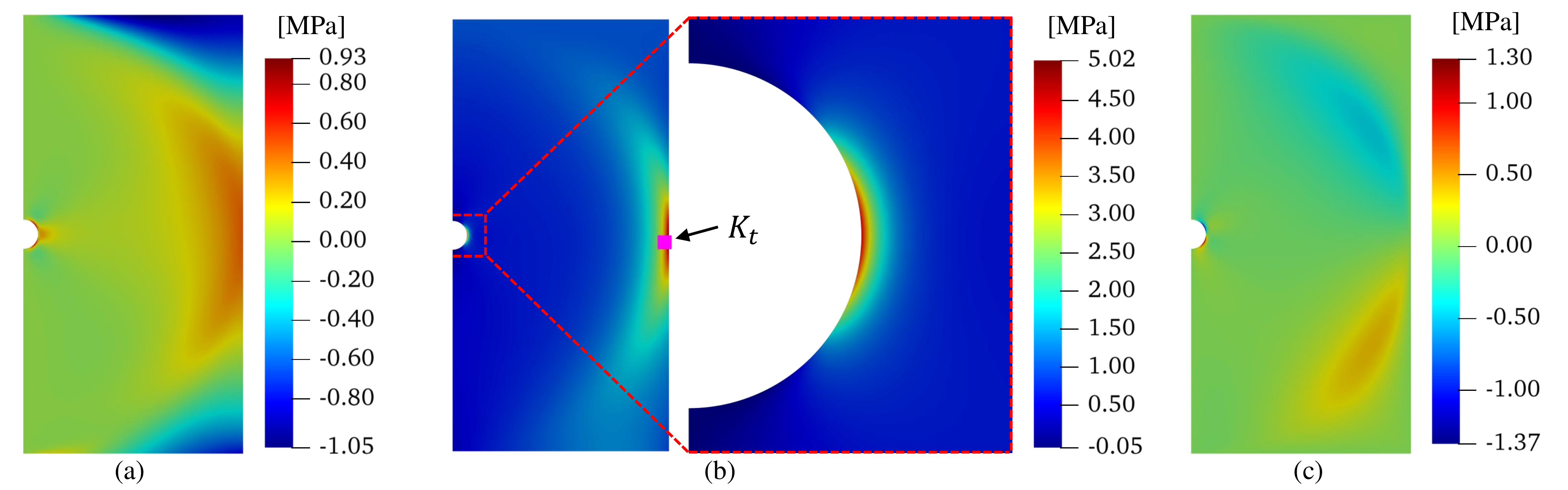}
	\centering
	\caption[Contours of the stresses (concentric fiber path following the semi-circular notch)]{Stress fields of composites reinforced by concentric fibers following the semi-circular notch: (a) stress in the $x$-direction $\sigma_{xx}$, (b) stress in the $y$-direction $\sigma_{yy}$ and the magnification around the semi-circular notch with the location of the stress concentration factor $K_t$ marked by the magenta square, (c) shear stress in the $x$-$y$ plane $\tau_{xy}$.}
	\label{concentric}
\end{figure}

\begin{figure}[!htbp]
    \includegraphics[scale=0.58]{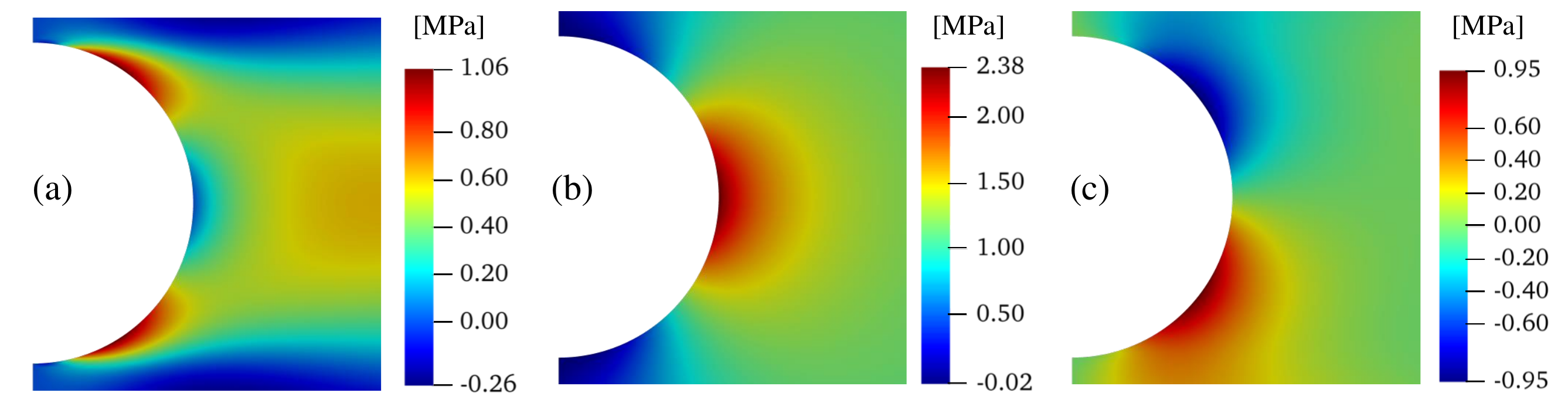}
	\centering
	\caption[Contours of the stresses (transverse straight fiber path)]{Stress fields of composites reinforced by transverse straight fibers: (a) stress in the $x$-direction $\sigma_{xx}$, (b) stress in the $y$-direction $\sigma_{yy}$, (c) shear stress in the $x$-$y$ plane $\tau_{xy}$.}
	\label{transverse}
\end{figure}

One of the fiber configurations investigated features concentric paths around the semi-circular notch (Fig \ref{paths}c). In this case, the fibers above and below the notch are transverse to the load and form relatively compliant regions. This can be clearly noted comparing the contour plots of the in-plane stress components for the concentric fiber paths (Fig. \ref{concentric}) to the cases of longitudinal (Fig. \ref{longitudinal}) and transverse (Fig. \ref{transverse}) fibers. As can be seen, the compliant regions slightly shield the notch and the maximum stress concentration factor of about $5.02$ is located at the middle point of the right-hand side edge of the plate. This is in contrast to the two linear fiber cases where the maximum stress is always located at the notch tip. The stress around the semi-circular notch is still higher than the nominal value ($T_y= 1$MPa) and the stress concentration factor $K_t$ there is roughly $4$.

Comparing the stress concentration to the linear fiber cases, it is interesting to note that the $K_t$ for the concentric path case is about $30\%$ lower of the longitudinal fiber case and about $111\%$ higher of the transverse fiber case. On the other hand, the overall plate stiffness is about $1/8$th of the longitudinal fiber case and about twice the transverse fiber case. Thus, in this configuration, curvilinear anisotropy indeed reduces the stress concentration compared to the longitudinal fiber case. However, this is possible only at the expense of the structural stiffness which is significantly lower than the longitudinal fiber case (although still higher than the transverse fiber case).

Among the linear fiber cases, the Tsai-Wu index takes the highest value for the transverse straight fiber path (Fig. \ref{paths}b): $0.050$. This is because the resin strength dominates in the loading direction. Although the stress concentration for the concentric fiber configuration is smaller than the one of the traditional longitudinal fiber composite, the maximum failure index $0.051$ is similar to the one of the transverse fiber configuration. Hence, this choice of curvilinear fiber path may not be so good for a damage tolerance due to the shear damage.


\subsubsection{Curvilinear fibers following the holomorphic paths}
\begin{figure}[!htbp]
    \includegraphics[scale=.58]{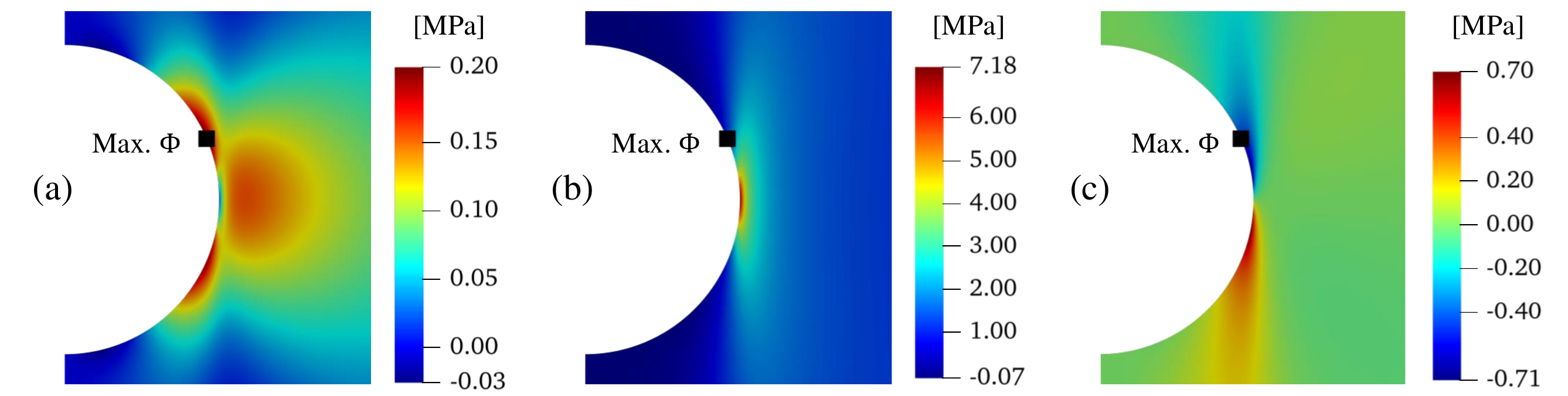}
	\centering
	\caption[Contours of the stresses (longitudinal straight fiber path)]{Stress fields of composites reinforced by longitudinal straight fibers with the maximum Tsai-Wu index $\Phi$ marked by the black square: (a) stress in the $x$-direction $\sigma_{xx}$, (b) stress in the $y$-direction $\sigma_{yy}$, (c) shear stress in the $x$-$y$ plane $\tau_{xy}$.}
	\label{longitudinal}
\end{figure}

\begin{figure}[!htbp]
    \includegraphics[scale=0.58]{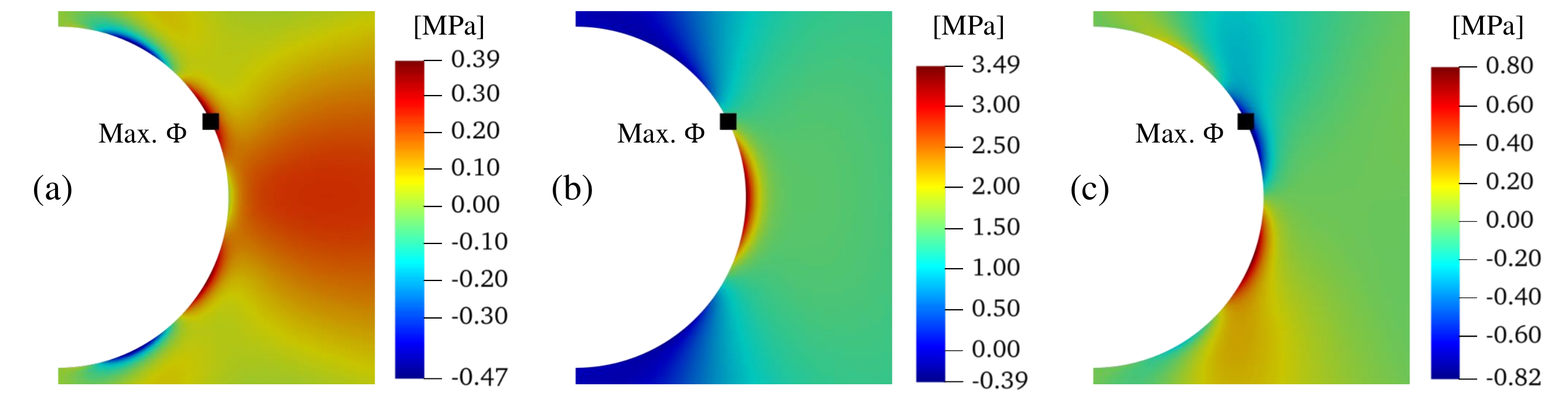}
	\centering
		\caption[Contours of the stresses (curvilinear fiber path following the holomorphic path)]{Stress fields of composites reinforced by curvilinear fibers following the holomorphic path with the maximum Tsai-Wu index $\Phi$ pointed marked by the black square: (a) stress in the $x$-direction $\sigma_{xx}$, (b) stress in the $y$-direction $\sigma_{yy}$, and (c) shear stress in the $x$-$y$ plane $\tau_{xy}$.}
    \label{holomorphic}
\end{figure}

Figure \ref{holomorphic} shows the stress contour plots of each in-plane stress component for the case of the plate reinforced by fibers following the holomorphic path described in \cite{salviato2019}. These contour plots look similar to the traditional longitudinal straight fiber composite (Fig. \ref{longitudinal}) with the maximum stress concentration located at the notch tip. However, the stress concentration factor $K_t$ is $3.49$, which is approximately half of the one of the longitudinal straight fiber composite. At the same time, the average stiffness $\bar{K}$ for the curvilinear fiber ply is 86.7 GN/mm, a value very close to the 88.2 GN/mm calculated for the longitudinal straight fiber case. This means that, although the stiffness is almost the same in both materials, the stress $\sigma_{yy}$ is significantly less concentrated around the semi-circular notch in the curvilinear fiber case than in the longitudinal straight fiber case. 

The maximum Tsai-Wu failure index for the curvilinear holomorphic fibers is $0.026$, which is much smaller than the ones from other fiber paths shown in Figs. \ref{paths}b, \ref{paths}c. However, the maximum Tsai-Wu index for the longitudinal straight fibers is way smaller, approximately six times smaller, but the stress concentration is significantly high. 
As can be seen in the Fig. \ref{longitudinal} and \ref{holomorphic}, the maximum Tsai-Wu indexes are located in the neighbourhood of the maximum stress of the x- and shear components for both cases. This indicates that the failures of the system could be experienced due to the resin and shear damage. 


\section{Optimization study}

After demonstrating the performance of the IGA formulation in simulating a set of different curvilinear fiber configurations and having investigated some of the effects of curvilinear anisotropy on the stress distributions, it is interesting to use our optimization framework to explore fiber configurations minimizing stress concentration and Tsai-Wu failure index. 

\subsection{Problem statement}
The optimization study defined in Eq. (\ref{opt1}) and (\ref{opt2}) is conducted in order to find optimal fiber paths which minimize the stress concentration factor and the Tsai-Wu failure index via SQP method on the elastic plate defined in Fig. \ref{plate} and Table \ref{eng_constants}. As initial condition, the parameters $C_{ij}$ of the level-set function describing the family of fiber paths, Eq. (\ref{level}), are chosen to take unit values.

\subsection{Optimal fiber path for the minimum stress concentration factor}
\begin{figure}[!htbp]
    \includegraphics[scale=.65]{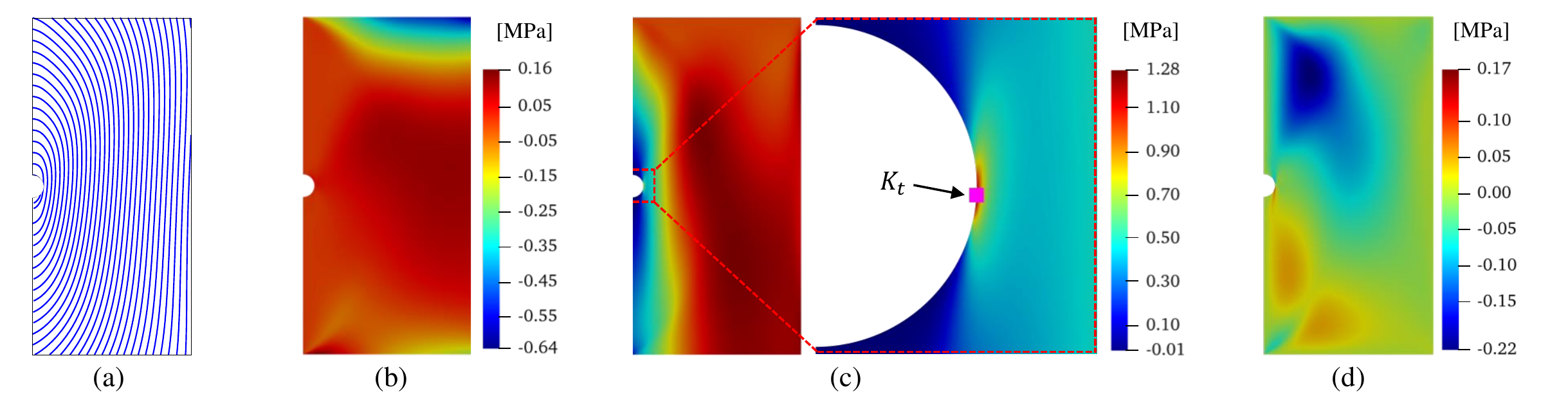}
	\centering
	\caption[Optimal fiber path for the minimum stress concentration factor and the contours of the stresses]{Optimal fiber path for the minimum stress concentration factor: (a), and the contour plots of: (b) the stress in the $x$-direction $\sigma_{xx}$, (c) the stress in the $y$-direction $\sigma_{yy}$ and the magnification around the semi-circular notch with the location of the stress concentration factor $K_t$ marked by the magenta square, (d) the shear stress in the $x-y$ plane $\tau_{xy}$.}
	\label{opt_concenration}
\end{figure}

 The optimal fiber paths identified via SQP are shown in Fig. \ref{opt_concenration}a, and the results in terms of stress concentration, failure index, and overall stiffness are summarized in Table \ref{results_2}. The contours of the stresses are shown in Fig. \ref{opt_concenration}b-d. The stress concentration for the optimized case 1.28 (Table \ref{results_1}a) which is remarkably 82 \% lower than the longitudinal fiber case and even lower than the isotropic case. This result implies that, by harnessing curvilinear anisotropy, it is possible to significantly reduce the severity of the semi-circular notch. 
 This cannot be achieved in isotropic materials unless the elastic properties are properly functionally-graded. 
 It is interesting to note that the obtained stress concentration is even lower than in the transverse straight fiber case, which features the lowest stress concentration among the configurations previously investigated. In fact, the stress concentration for the optimal fiber path case (Table \ref{results_1}a) is lower by 46 \%. 
 In conclusion, harnessing curvilinear anisotropy can significantly reduce the stress concentration even compared to isotropic case. Factoring the increase in strength and stiffness provided by the embedded fibers, it is clear that media featuring CTI can surpass the structural performance of traditional composites. Future work will focus on translating these results to other notch configurations including e.g. sharp V-notches or U-notches.

\subsection{Optimal fiber path for the minimum Tsai-Wu failure index}
The optimization study for the minimum Tsai-Wu index is also implemented under the same condition for the minimum stress concentration. The initial design variables $C_{ij}$ are decided as the longitudinal straight fiber path shown in Fig. \ref{paths}a such that:
\begin{equation}
	C_{ij} = \left\{
	\begin{array}{rl}
	1 & \text{if } (i,j) = (1,0), \\
	0 & \text{otherwise}
	\end{array} \right.
\end{equation}
because this fiber path provides the minimum Tsai-Wu failure index based on the four different fiber configurations (Fig. \ref{paths}a and Table \ref{results_1}).

As a result of the optimization, the design variables remained as same as the initial conditions indicating that the longitudinal straight fiber case is the one minimizing the failure index (Table \ref{results_1}). In terms of Tsai-Wu index, conventional longitudinal straight fiber path is the best option, however, the stress concentration is very significant compared to the optimal fiber case of the minimum stress concentration. Thus, it is interesting to make a comparison between the two focusing on Tsai-Wu index as discussed next.

\subsection{Damage initiation analysis}
\begin{figure}[!htbp]
	\includegraphics[scale=.7]{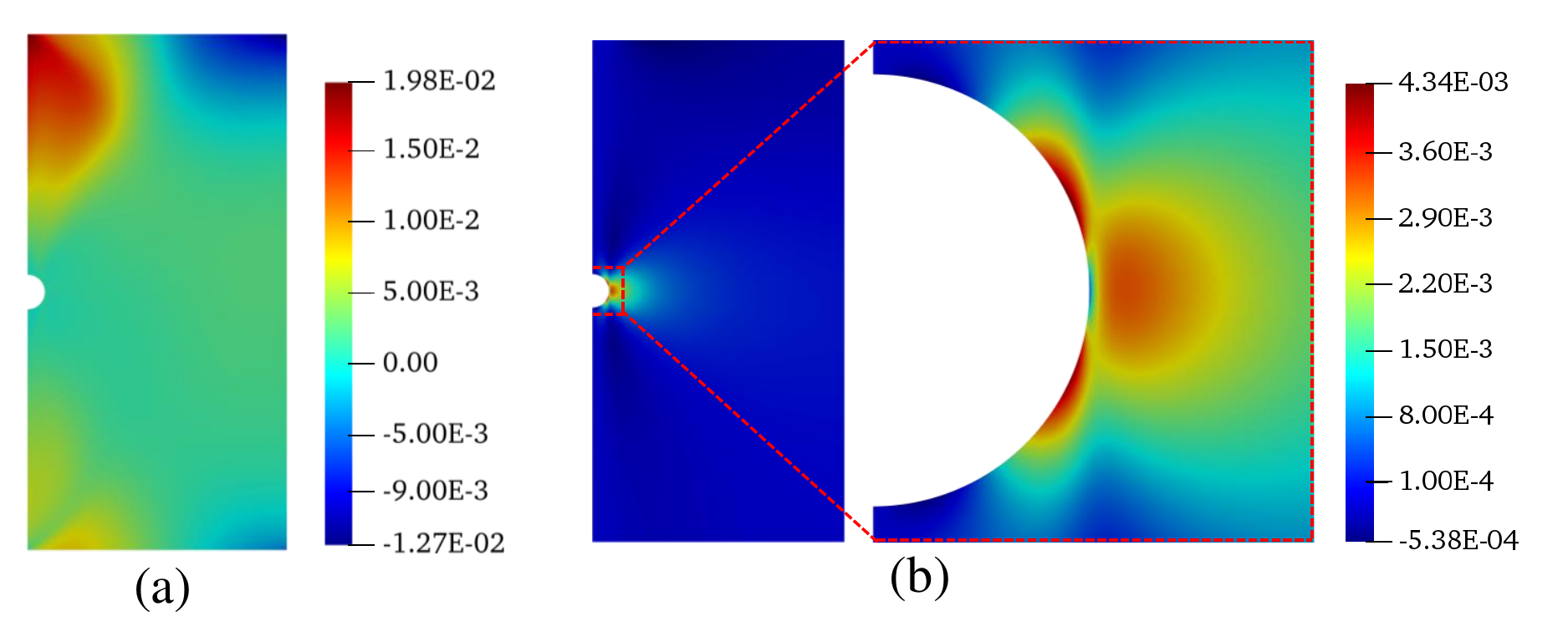}
	\centering
	\caption[Contours of the Tsai-Wu failure index of the optimal fiber paths with the radius of $10$ mm]{Contour plots of the Tsai-Wu failure index of the optimal fiber paths: (a) for the minimum stress concentration factor, and (b) for the minimum Tsai-Wu failure index and the magnification around the semi-circular notch.}
	\label{T-W}
\end{figure}

Based on Table \ref{results_2}, the maximum Tsai-Wu failure index for the fiber path that provides the minimum stress concentration is found as $0.020$ and it is located at the very top-left node. This value is about five times higher than the one from the longitudinal straight fiber path, which is the optimal fiber path for minimum Tsai-Wu index. Table \ref{dominance} includes the local stresses of the each component, and the local dominance of the ply strength at the points of the maximum Tsai-Wu failure index. The local ratios between stresses to the lamina strength such that:\\
For the longitudinal fiber direction:
\begin{equation}\label{dominance_1}
	\sigma_1/\sigma_{1}^f = \left\{
	\begin{array}{rl}
		\sigma_1/\sigma_{1t}^f & \text{if } \sigma_1 \geq 0 \\
		\sigma_1/\sigma_{1c}^f & \text{if } \sigma_1 < 0 \\
	\end{array} \right.
\end{equation}
For the transverse fiber direction:
\begin{equation}\label{dominance_2}
	\sigma_2/\sigma_{2}^f = \left\{
	\begin{array}{rl}
	\sigma_2/\sigma_{2t}^f & \text{if } \sigma_2 \geq 0 \\
	\sigma_2/\sigma_{2c}^f & \text{if } \sigma_2 < 0 \\
	\end{array} \right.
\end{equation}
while for the shear component the ratio, $\tau_{12}/\tau_{12}^f$  is independent of the sign of the local shear stress.
Considering the case of the optimal fibers for the minimum stress concentration (Fig. \ref{opt_concenration}a), the transverse stress component is the closest to the related strength value, showing that matrix failure by e.g. splitting at the very top-left of the plate (Fig. \ref{T-W}a) is the most likely damage mechanism. However, the resulting crack would probably grow along the curvilinear fiber path and eventually be arrested. Thus, even though the plate begins to experience the failure, it might not immediately be exposed to be an ultimate failure. 
In the medium optimized for minimum Tsai-Wu index, the transverse and the shear components of the strengths take the highest values compared to the respective strengths, which implies that splitting crack is again the dominant failure mode. However, as can be seen in Fig. \ref{T-W}b, the splitting crack will initiate close to the notch tip 
and propagate undisturbed following the longitudinal fiber path until reaching the end of the plate. 

Thus, it is very difficult to predict which fiber configuration will attain the highest load capacity. Of course, the Tsai-Wu failure criterion is not equipped to predict stable crack growth and quasibrittle fracture mechanics models will be used in future works to investigate the highly nonlinear progressive damage following initiation \cite{okabe2020,Salviato2016,Kirane12016,Kirane2016,Qiao2020}.

\section{Conclusion}
In this work we explored an Iso-Geometric Analysis (IGA) computational framework for the simulation of media featuring curvilinear anisotropy. Based on our results we can formulate the following conclusions on its numerical performance:

\begin{enumerate}
    \item thanks to the exact geometric representation and the enriched continuity between elements, NURBS-based IGA outperforms classical FEA in terms of computational efficiency, run time, and accuracy in the description of field variables for the same number of degrees-of-freedom; 
    \item For the configurations investigated in this work, the use of IGA resulted in an average reduction of the total simulation time of about $40\%$ compared to FEM given the same error in the estimate of the maximum stress 
    and solution scheme of the reduced system of equations;
    \item the higher efficiency of IGA makes it an interesting alternative to FEM for the simulation of large and complex structures made of materials featuring curvilinear anisotropy. Further, this characteristic makes it particularly suitable for optimization studies which generally require several simulations before reaching the optimal solution;
\end{enumerate}

\begin{figure}[!htbp]
	\includegraphics[scale=0.7]{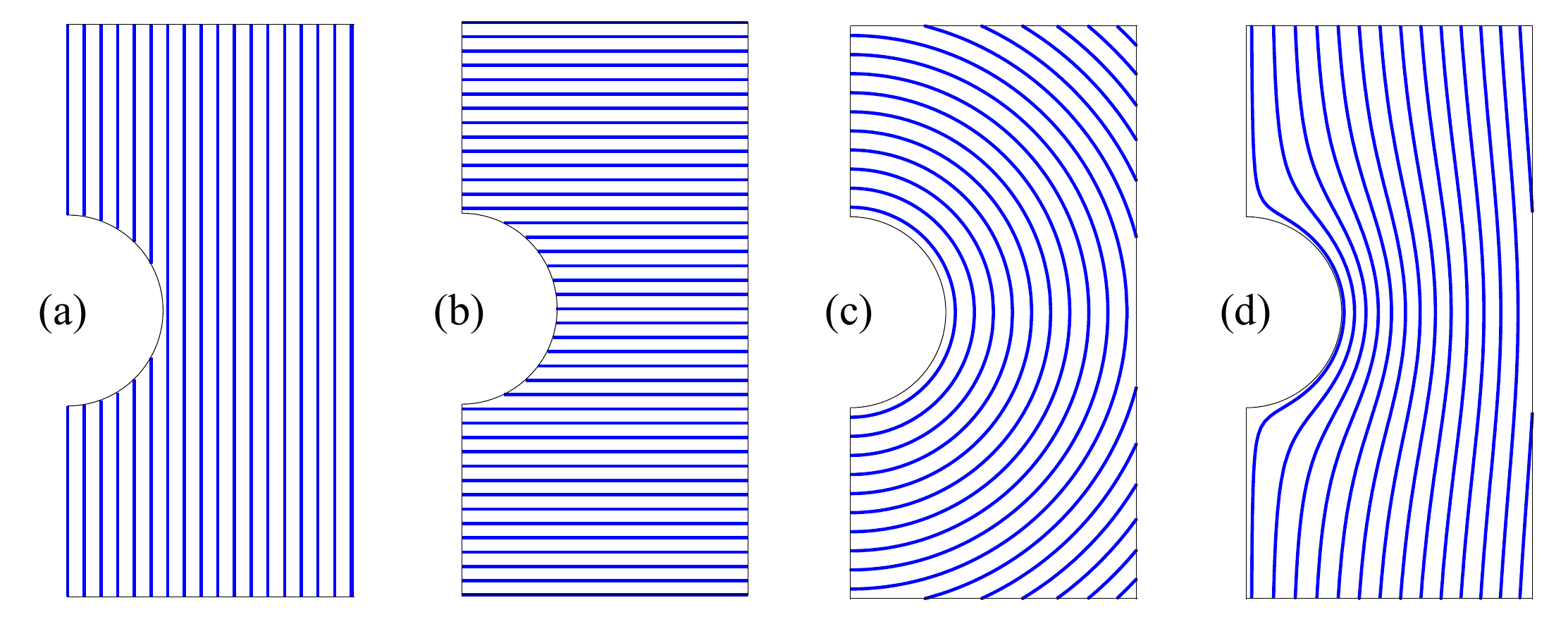}
	\centering
	\caption[Composites reinforced by four different curvilinear fibers]{Composites reinforced by different fiber configurations: (a) Longitudinal straight fiber paths, (b) Transverse straight fiber paths, (c) Concentric fiber paths following the semi-circular notch, (d) Curvilinear fiber paths following the semi-circular notch.}
	\label{paths}
\end{figure}

We demonstrated the use of the IGA framework using a plate weakened by a semi-circular notch with the four fiber configurations shown in Fig. \ref{paths} as a case study. Then, we also performed an optimization study aimed at identifying the paths leading to the lowest stress concentrations or values of the Tsai-Wu failure index. Thanks to these studies, we can draw the following conclusions on the mechanical performance of plates featuring curvilinear anisotropy: 

\begin{enumerate}\addtocounter{enumi}{3}
	\item as expected, curvilinear anisotropy has a significant effect on the elastic stress and strain fields as well as the overall stiffness of the plate;
	
	\item in curvilinear anisotropic media, the regions of high stress concentration can be widely distributed over the domain. Hence, when performing optimization studies, it is important to utilize a uniform refinement instead of a local refinement around the notch. Guaranteeing a uniform accuracy throughout the domain may become very computational expensive with FEM. In contrast, IGA significantly mitigates this problem thanks to the higher accuracy and efficiency;
	
	
	\item by optimizing the fiber paths in the plate, we were able to reduce the stress concentration factor, $K_t$, to 1.28 with the maximum stress located at the notch tip. This represents a reduction of about $82\%$ compared to the classical longitudinal linear fiber case and a reduction of 58\% compared to the isotropic case;
	
	\item on the other hand, we showed that the longitudinal linear fiber path provides the minimum Tsai-Wu failure index, with damage initiation in the form of a mode II splitting crack close to the tip of the notch. 
	In contrast, the path minimizing the stress concentration gives a failure index about five times larger, with damage initiation as mode I splitting crack. However, this result does not allow significant conclusions on the ultimate capacity of the plates since this depends on the progressive damage evolution up to failure. It is likely that, after initiation, the splitting crack in the curvilinear anisotropic case will be significantly slowed down and arrested by the fibers before final failure. On the other hand, the splitting crack in the longitudinal fiber case will propagate unstably until reaching the top and bottom boundaries of the plate. To investigate damage progression in curvilinear anisotropic media, future works will focus on the use of quasibrittle fracture mechanics computational models \cite{okabe2020,Salviato2016,Kirane12016,Kirane2016,Qiao2020}. These are quintessential to capture the formation of large Fracture Process Zones (FPZs) and distributed damage in composites which are a significant source of size effects \cite{salviato1,salviato2,Qiao2020};
	
	\item another interesting result is that the curvilinear fiber paths designed to optimize the electric conductivity of the plate~\cite{salviato2019} showed also an outstanding mechanical performance. In fact, the stress concentration factor was found to be 3.49, about 51\% lower than the longitudinal linear fiber case with similar levels of plate stiffness. This result is particularly promising because it shows that it is possible to harness curvilinear anisotropy not only to obtain superior mechanical performance compared to traditional composites, but also to introduce novel multi-functional properties. Future works will focus on the formulation of novel multi-objective optimization schemes within the IGA framework to take advantage of this new design space.

	

\end{enumerate}

\clearpage

\linespread{1}\selectfont
\bibliographystyle{elsarticle-num}
\bibliography{bib}

\begin{thebibliography}{10}
\expandafter\ifx\csname url\endcsname\relax
  \def\url#1{\texttt{#1}}\fi
\expandafter\ifx\csname urlprefix\endcsname\relax\def\urlprefix{URL }\fi
\expandafter\ifx\csname href\endcsname\relax
  \def\href#1#2{#2} \def\path#1{#1}\fi

\bibitem{Das2019PreparationDO}
T.~K. Das, P.~Ghosh, N.~C. Das, {Preparation, development, outcomes, and
  application versatility of carbon fiber-based polymer composites: a review},
  Advanced Composites and Hybrid Materials 2~(2) (2019) 214--233.

\bibitem{Harris2002}
C.~E. Harris, J.~H. Starnes, M.~J. Shuart, {Design and manufacturing of
  aerospace composite structures, state-of-the-art assessment}, Journal of
  Aircraft 39~(4) (2002) 545--560.

\bibitem{Othman2018}
R.~Othman, N.~I. Ismail, M.~A. A.~H. Pahmi, M.~H.~M. Basri, H.~Sharudin, A.~R.
  Hemdi, {Application of Carbon Fiber Reinforced Plastics in Automotive
  Industry: a Review}, Journal of Mechanical Manufacturing 1~(June) (2019)
  144--154.

\bibitem{dirk2012engineering}
H.-J.~L. Dirk, C.~Ward, K.~D. Potter, The engineering aspects of automated
  prepreg layup: History, present and future, Composites Part B: Engineering
  43~(3) (2012) 997--1009.

\bibitem{CROFT2011484}
K.~Croft, L.~Lessard, D.~Pasini, M.~Hojjati, J.~Chen, A.~Yousefpour,
  Experimental study of the effect of automated fiber placement induced defects
  on performance of composite laminates, Composites Part A: Applied Science and
  Manufacturing 42~(5) (2011) 484 -- 491.

\bibitem{GAO201565}
W.~Gao, Y.~Zhang, D.~Ramanujan, K.~Ramani, Y.~Chen, C.~B. Williams, C.~C. Wang,
  Y.~C. Shin, S.~Zhang, P.~D. Zavattieri, The status, challenges, and future of
  additive manufacturing in engineering, Computer-Aided Design 69 (2015) 65 --
  89.

\bibitem{Mansfield1953}
E.~H. Mansfield, {Neutral holes in plane sheet: Reinforced holes which are
  elastically equivalent to the uncut sheet}, Quarterly Journal of Mechanics
  and Applied Mathematics 6~(3) (1953) 370--378.

\bibitem{Hyer1987}
M.~Hyer, R.~Charette, {Innovative Design of Composite Structures: Use of
  Curvilinear Fiber Format to Improve Structural Efficiency}, Tech. rep.,
  University of Maryland (1987).

\bibitem{Hyer1991}
M.~W. Hyer, H.~H. Lee, {The use of curvilinear fiber format to improve buckling
  resistance of composite plates with central circular holes}, Composite
  Structures 18~(3) (1991) 239--261.

\bibitem{Gurdal2008}
Z.~G{\"{u}}rdal, B.~F. Tatting, C.~K. Wu, {Variable stiffness composite panels:
  Effects of stiffness variation on the in-plane and buckling response},
  Composites Part A: Applied Science and Manufacturing 39~(5) (2008) 911--922.

\bibitem{Cho2007}
H.~K. Cho, R.~E. Rowlands, {Reducing tensile stress concentration in perforated
  hybrid laminate by genetic algorithm}, Composites Science and Technology
  67~(13) (2007) 2877--2883.

\bibitem{Cho2009}
H.~K. Cho, R.~E. Rowlands, {Optimizing fiber direction in perforated
  orthotropic media to reduce stress concentration}, Journal of Composite
  Materials 43~(10) (2009) 1177--1198.

\bibitem{HondaS.2010Mosc}
S.~Honda, K.~Owatari, Y.~Narita, {Minimization of stress concentration for
  laminated composite plates with curvilinearly shaped fibers}, Japan Society
  of Mechanical Engineers, Part A 76~(769) (2010) 1139--1146.

\bibitem{honda2013multi}
S.~Honda, T.~Igarashi, Y.~Narita, Multi-objective optimization of curvilinear
  fiber shapes for laminated composite plates by using nsga-ii, Composites Part
  B: Engineering 45~(1) (2013) 1071--1078.

\bibitem{Vijayachandran12020}
A.~A. Vijayachandran, P.~Davidson, A.~M. Waas, {Optimal fiber paths for
  robotically manufactured composite structural panels}, International Journal
  of Non-Linear Mechanics 126~(July) (2020) 103567.

\bibitem{Vijayachandran22020}
A.~A. Vijayachandran, P.~Davidson, A.~M. Waas, {Optimal steered fiber paths for
  maximizing biaxial buckling load of a flat plate manufactured using AFP}, in:
  AIAA Scitech 2020 Forum, 2020.

\bibitem{mat16}
R.~Matsuzaki, M.~Ueda, M.~Namiki, T.-K. Jeong, H.~Asahara, K.~Horiguchi,
  T.~Nakamura, A.~Todoroki, Y.~Hirano, Three-dimensional printing of
  continuous-fiber composites by in-nozzle impregnation, Scientific reports 6
  (2016) 23058.

\bibitem{Mat20}
F.~Van Der~Klift, Y.~Koga, A.~Todoroki, M.~Ueda, Y.~Hirano, R.~Matsuzaki,
  et~al., 3d printing of continuous carbon fibre reinforced thermo-plastic
  (cfrtp) tensile test specimens, Open Journal of Composite Materials 6~(01)
  (2016) 18.

\bibitem{salviato2019}
M.~Salviato, S.~E~Phenisee, Enhancing the electrical and thermal conductivities
  of polymer composites via curvilinear fibers: An analytical study,
  Mathematics and Mechanics of Solids (03 2019).

\bibitem{hughes2005isogeometric}
T.~J. Hughes, J.~A. Cottrell, Y.~Bazilevs, Isogeometric analysis: Cad, finite
  elements, nurbs, exact geometry and mesh refinement, Computer methods in
  applied mechanics and engineering 194~(39-41) (2005) 4135--4195.

\bibitem{nguyen2015isogeometric}
V.~P. Nguyen, C.~Anitescu, S.~P. Bordas, T.~Rabczuk, Isogeometric analysis: an
  overview and computer implementation aspects, Mathematics and Computers in
  Simulation 117 (2015) 89--116.

\bibitem{cottrell2009isogeometric}
J.~A. Cottrell, T.~J. Hughes, Y.~Bazilevs, Isogeometric analysis: toward
  integration of CAD and FEA, John Wiley \& Sons, 2009.

\bibitem{Bordas3}
C.~H. Thai, A.~Ferreira, S.~P.~A. Bordas, T.~Rabczuk, H.~Nguyen-Xuan,
  Isogeometric analysis of laminated composite and sandwich plates using a new
  inverse trigonometric shear deformation theory, European Journal of
  Mechanics-A/Solids 43 (2014) 89--108.

\bibitem{locking}
L.~B. Da~Veiga, A.~Buffa, C.~Lovadina, M.~Martinelli, G.~Sangalli, An
  isogeometric method for the reissner--mindlin plate bending problem, Computer
  Methods in Applied Mechanics and Engineering 209 (2012) 45--53.

\bibitem{Bordas2}
N.~Valizadeh, S.~Natarajan, O.~A. Gonzalez-Estrada, T.~Rabczuk, T.~Q. Bui,
  S.~P. Bordas, Nurbs-based finite element analysis of functionally graded
  plates: static bending, vibration, buckling and flutter, Composite Structures
  99 (2013) 309--326.

\bibitem{Logan2016}
D.~L. Logan, {A first course in the finite element method}, 6th Edition,
  Cengage Learing, Boston, 2016.

\bibitem{gord74}
W.~J. Gordon, R.~F. Riesenfeld, B-spline curves and surfaces, in: R.~E.
  Barnhill, R.~F. Riesenfeld (Eds.), Computer Aided Geometric Design, Academic
  Press, New York, 1974, pp. 95--126.

\bibitem{Riesenfeld1973}
R.~F. Riesenfeld, {Applications of B-Spline Approximation to Geometric Problems
  of Computer-aided Design}, Ph.D. thesis, Syracuse University (1973).

\bibitem{Cox72}
M.~G. Cox, {The Numerical Evaluation of B-Splines*}, IMA Journal of Applied
  Mathematics 10~(2) (1972) 134--149.

\bibitem{DEBoor1972}
C.~{De Boor}, {On Calculating with B-Splines}, Journal of Approximation Theory
  6 (1972) 50--62.

\bibitem{de1978practical}
C.~De~Boor, A practical guide to splines, Vol.~27, springer-verlag New York,
  1978.

\bibitem{Vers75}
K.~Versprille, Computer-aided design applications of the rational b-spline
  approximation form, Ph.D. thesis, Syracuse University (1975).

\bibitem{tiller83}
W.~Tiller, Rational b-splines for curve and surface representation, IEEE
  Computer Graphics and Applications 3~(6) (1983) 61--69.

\bibitem{Piegl1991}
L.~Piegl, {On NURBS: A Survey}, IEEE Computer Graphics {\&} Applicntions 10~(1)
  (1991) 55--71.

\bibitem{piegl1997nurbs}
L.~Piegl, W.~Tiller, The NURBS book, Springer-Verlag, 1997.

\bibitem{de1972calculating}
C.~De~Boor, On calculating with b-splines, Journal of Approximation Theory
  6~(1) (1972) 50--62.

\bibitem{Mortenson1985}
M.~E. Mortenson, {Geometric Modeling}, John Wiley and Sons, Ltd., New York,
  1985.

\bibitem{Farin1993}
G.~Farin, {Curves and Surfaces for Computer-Aided Geometric Design- A Practical
  Guide}, 4th Edition, Academic Press, INC., Boston, 1993.

\bibitem{Rogers1986}
D.~F. Rogers, J.~A. Adams, {Mathematical Elements for Computer Graphics},
  McGraw-Hill, New York, 1986.

\bibitem{Gibson1994}
R.~F. Gibson, {Principles of Composite Material Mechanics}, McGraw-Hill, New
  York, 1994.

\bibitem{kol03}
L.~P. Kollár, G.~S. Springer, Mechanics of Composite Structures, Cambridge
  University Press, New York, 2003.

\bibitem{Suzuki2019}
K.~Suzuki, {Isogeometric Computational Modeling of Curvilinear Fiber
  Composites}, Master's thesis, University of Washington (2019).

\bibitem{Arora2012}
J.~S. Arora, {Introduction to Optimum Design}, 3rd Edition, Elsevier Inc.,
  2012.

\bibitem{Nocedal2006}
J.~Nocedal, S.~J. Wright, {Numerical Optimization}, 2nd Edition, Springer,
  2006.

\bibitem{karush39}
W.~Karush, Minima of functions of several variables with inequalities as side
  conditions., Master's thesis, University of Chicago, Department of
  Mathematics (1939).

\bibitem{Kuhn1951}
H.~W. Kuhn, A.~W. Tucker, {Nonlinear programming}, in: Proceedings of 2nd
  Berkeley Symposium, University of California Press, Berkeley, 1951, pp.
  481--492.

\bibitem{Broyden1970}
C.~G. Broyden, {The convergence of a class of double-rank minimization
  algorithms}, IMA Journal of Applied Mathematics 6~(1) (1970) 76--90.

\bibitem{Fletcher1970}
R.~Fletcher, {A new approach to variable metric algorithms}, The Computer
  Journal 13~(3) (1970) 317--322.

\bibitem{Goldfarb1970}
D.~Goldfarb, {A Family of Variable-Metric Methods Derived by Variational
  Means}, Mathematics of Computation 24~(109) (1970) 23.

\bibitem{Shanno1970}
D.~F. Shanno, {Conditioning of Quasi-Newton Methods for Function Minimization},
  Mathematics of Computation 24~(111) (1970) 647--656.

\bibitem{Mathworks2020}
{Optimization Toolbox™ User 's Guide}, The MathWorks, Inc., Natick,
  Massachusetts, United States, 2020.

\bibitem{Tsai1971}
S.~W. Tsai, E.~M. Wu, {A General Theory of Strength for Anisotropic Materials},
  Journal of Composite Materials 5~(1) (1971) 58--80.

\bibitem{bazant}
Z.~Ba{\v{z}}ant, Scaling theory for quasibrittle structural failure,
  Proceedings of the National Academy of Sciences 101~(37) (2004) 13400--13407.

\bibitem{Salviato2016}
M.~Salviato, S.~E. Ashari, G.~Cusatis, {Spectral stiffness microplane model for
  damage and fracture of textile composites}, Composite Structures 137 (2016)
  170--184.

\bibitem{salviato1}
M.~Salviato, K.~Kirane, S.~Ashari, Z.~Ba{\v{z}}ant, G.~Cusatis, Experimental
  and numerical investigation of intra-laminar energy dissipation and size
  effect in two-dimensional textile composites, Composites Science and
  Technology 135 (2016) 67--75.

\bibitem{salviato2}
M.~Salviato, K.~Kirane, Z.~Ba{\v{z}}ant, G.~Cusatis, Mode {I} and {II}
  interlaminar fracture in laminated composites: a size effect study, Journal
  of Applied Mechanics 86~(9) (2019).

\bibitem{okabe2020}
Y.~Kumagai, S.~Onodera, M.~Salviato, T.~Okabe, {Multiscale analysis and
  experimental validation of crack initiation in quasi-isotropic laminates},
  International Journal of Solids and Structures 193-194 (2020) 172--191.

\bibitem{Kirane12016}
K.~Kirane, M.~Salviato, Z.~P. Ba{\v{z}}ant, {Microplane triad model for simple
  and accurate prediction of orthotropic elastic constants of woven fabric
  composites}, Journal of Composite Materials 50~(9) (2016) 1247--1260.

\bibitem{Kirane2016}
K.~Kirane, M.~Salviato, Z.~P. Bazant, {Microplane-Triad Model for Elastic and
  Fracturing Behavior of Woven Composites}, Journal of Applied Mechanics,
  Transactions ASME 83~(4) (2016) 1--14.

\bibitem{Qiao2020}
Y.~Qiao, Q.~Zhang, M.~Salviato, {Effects of In-situ Stress State on the Plastic
  Deformation, Fracture, and Size Scaling of Thermoset Polymers and Related
  Fiber-reinforced Composites}, in: ASC 35th Technical Conference, no. July,
  2020.

\end{thebibliography}

\clearpage

\begin{table}[!htbp]
    \caption[Table of Elastic Material Properties]{Elastic material properties}
	\centering
	\begin{tabular}{cccc}
		\hline
		Symbol & & Unit & Property \\
		\hline
		$E_1$ & 181 & GPa & Longitudinal modulus\\
		$E_2$ & 10.3 & GPa & Transverse modulus\\
		$G_{12}$ & 7.17 & GPa & Shear modulus\\
		$\nu_{12}$ & 0.28 & - & Poisson's ratio\\
		$\sigma_{1t}^f$ & 1500 & MPa & Longitudinal strength in tension\\
		$\sigma_{1c}^f$ & 1500 & MPa & Longitudinal strength in compression\\
		$\sigma_{2t}^f$ & 40 & MPa & Transverse strength in tension\\
		$\sigma_{2c}^f$ & 246 & MPa & Transverse strength in compression \\
		$\sigma_{12}^f$ & 68 & MPa & Shear strength\\
	    \hline
	\end{tabular}
	\label{eng_constants}
\end{table}

\begin{table}[!htbp]
	\caption[Summery of the stress concentration factor, the average stiffness, and the Tsai-Wu index of a variety of CTI composites]{Table of the stress concentration factor $K_t$, the average stiffness on the top edge $\bar{K}$, and the maximum Tsai-Wu failure index $\Phi$ of: (a) Longitudinal straight fiber path (Fig. \ref{plate}a), (b) Transverse straight fiber path (Fig. \ref{plate}b), (c) Concentric fiber path (Fig. \ref{plate}c), (d) Curvilinear fiber path following the semi-circular notch (Fig. \ref{plate}d).}
	\centering
	\begin{tabular}{cccccc}
		\hline
		& (a) & (b) & (c) & (d) \\
		\hline
		$K_t$ & $7.18$ & $2.38$ & $5.02$ & $3.49$  \\
		$\bar{K}$ [GN/mm]    & $88.20$ & $5.11$ & $10.78$ & $86.72$  \\
		$\Phi$ & $0.004$ & $0.050$ & $0.051$ & $0.026$  \\
		\hline
	\end{tabular}
	\label{results_1}
\end{table}

\begin{table}[!htbp]
	\caption[Summery of the stress concentration factor, the Tsai-Wu failure index, and the average stiffness of the optimal fiber paths]{Table of the stress concentration factor $K_t$, the average stiffness on the top edge $\bar{K}$, and the Maximum Tsai-Wu failure index $\Phi$ of:  (1) optimal fiber path for the minimum $K_t$, (2) the optimal fiber path for the minimum $\Phi$.}
	\centering
	\begin{tabular}{ccc}
		\hline
		& (1) & (2) \\
		\hline
		$K_t$ & $1.28$ & $7.18$ \\
		$\bar{K}$  [GN/mm] & $33.38$ & $88.20$  \\
		$\Phi$ & $0.020$ & $0.004$ \\
		\hline
	\end{tabular}
	\label{results_2}
\end{table}

\begin{table}[!htbp]
	\caption[The local stresses and the local stress dominance of ply strengths]{Table of the local stresses and the local stress dominance of ply strengths at the points of the maximum Tsai-Wu failure index for: (1) optimal fiber path for the minimum $K_t$, (2) the optimal fiber path for the minimum Tsai-Wu failure index $\Phi$.}
	\centering
	\begin{tabular}{ccccccc}
		\hline
		& $\sigma_1$ [MPa] & $\sigma_2$ [MPa] & $\tau_{12}$ [MPa] & $\sigma_1/\sigma_1^f$ & $\sigma_2/\sigma_2^f$ & $\tau_{12}/\tau_{12}^f$ \\
		\hline
		(1) & $0.061$ & $0.941$ & $0.238$ & $4.0E-5$ & $2.4E-2$ & $3.5E-3$ \\
		(2) & $1.112$ & $0.205$ & $0.480$ & $7.4E-4$ & $5.1E-3$ & $7.1E-3$ \\
		\hline
	\end{tabular}
	\label{dominance}
\end{table}

\end{document}